\documentclass[a4paper,fleqn,usenatbib]{mnras}
\usepackage[T1]{fontenc}
\usepackage{ae,aecompl}
\usepackage{amsmath}
\usepackage{amsfonts}
\usepackage{amssymb}
\usepackage{graphicx}
\usepackage{color,xcolor}
\def\mnras{MNRAS}
\def\aap{A$\&$A}
\def\nat{Nature}
\def\apj{ApJ}
\def\apjl{ApJ Letters}
\def\apjs{ApJS}
\def\pasp{PASP}
\def\aj{AJ}

\title[Proplyds in the Flame Nebula NGC 2024]{Proplyds in the Flame Nebula NGC 2024}
\author[T. J. Haworth et al. ]{Thomas J. Haworth$^1$\thanks{t.haworth@qmul.ac.uk}, Jinyoung S. Kim$^2$, Andrew J. Winter$^3$, Dean C. Hines$^4$, \newauthor Cathie J. Clarke$^5$, Andrew D. Sellek$^5$, Giulia Ballabio$^{1,6}$, Karl R. Stapelfeldt$^7$ \\
$^1$Astronomy Unit, School of Physics and Astronomy, Queen Mary University of London, Mile End Road, London, UK \\
$^2$Steward Observatory, University of Arizona, 933 N. Cherry Ave, Tucson, AZ 85721-0065, USA \\
$^3$Institut f\"{u}r Theoretische Astrophysik, Zentrum f\"{u}r Astronomie, Heidelberg University, Albert Ueberle Str. 2, 69120 Heidelberg,\\  Germany\\
$^4$Space Telescope Science Institute, 3700 San Martin Drive, Baltimore, MD 21218 USA \\
$^5$Institute of Astronomy, University of Cambridge, Madingley Road, Cambridge, CB3 0HA, UK\\
$^6$School of Physics and Astronomy, University of Leicester, Leicester, LE1 7RH, UK\\
$^7$Jet Propulsion Laboratory, California Institute of Technology, 4800 Oak Grove drive, Pasadena CA 91109, USA} 
\date{}
\begin{document}
\maketitle
\begin{abstract}
{A recent survey of the inner $0.35\times0.35$\,pc of the NGC 2024 star forming region revealed two distinct millimetre continuum disc populations that appear to be spatially segregated by the boundary of a dense cloud}. The eastern (and more embedded) population is $\sim0.2-0.5$\,Myr old, with {an ALMA mm continuum disc detection rate of about $45\,$per cent}. {However this drops to only $\sim15\,$per cent in the 1\,Myr western population}. When presenting this result, \cite{vanTerwisga20} suggested that the two main UV sources, IRS 1 (a B0.5V star in the western region) and IRS 2b (an O8V star in the eastern region, but embedded) have \textit{both} been evaporating the discs in the depleted western population. 

In this paper we report the firm discovery in archival HST data of 4 proplyds and 4 further candidate proplyds in NGC 2024, confirming that external photoevaporation of discs is occurring. However, the locations of these proplyds changes the picture. Only three of them are in the {depleted western} population and their evaporation is dominated by IRS 1, with no obvious impact from IRS 2b. The other 5 proplyds are in the younger eastern region and being evaporated by IRS 2b. We propose that both populations are subject to significant external photoevaporation, {which happens} throughout the region wherever discs are not sufficiently shielded by the interstellar medium. {The external photoevaporation and severe depletion of mm grains in the 0.2-0.5\,Myr eastern part of NGC 2024 would be in competition even with very early planet formation.}

\end{abstract}

\section{Introduction}
\label{sec:intro}

Planet formation is increasingly being considered in the wider context of star forming regions \citep[for a recent review see][]{2020arXiv200707890P}. The environmental impact on discs and planets from the surroundings may take place through external photoevaporation \citep[e.g.][]{1999ApJ...515..669S, 2004ApJ...611..360A, 2016MNRAS.457.3593F, 2016arXiv160501773G, Kim16, 2017AJ....153..240A, 2018ApJ...860...77E, 2019MNRAS.485.3895H, 2019MNRAS.485.4893N, 2019MNRAS.486.4354R, 2019MNRAS.490.5678C, 2020MNRAS.491..903W}, gravitational encounters \citep[e.g.][]{2006A&A...452..897C, 2015MNRAS.449.1996D, 2018ApJ...859..150R, Winter18c, 2019MNRAS.482..732C, 2020arXiv200715666M}, chemical enrichment, {such as through short lived radionuclides from supernovae} \citep[][]{2016ApJ...826...22K, 2017MNRAS.469.1117C, 2019NatAs...3..307L} and cosmic ray ionisation \citep{2020arXiv200600019K}. In terms of direct impact on the physical characteristics (mass/radius) of an established disc, external photoevaporation is generally thought to dominate over encounters \citep{2001MNRAS.325..449S, Winter18b}. Cosmic rays may affect the initial conditions of discs through the strength of magnetic braking \citep{2020arXiv200600019K}. 

\begin{figure}
    \centering
    \includegraphics[width=\columnwidth]{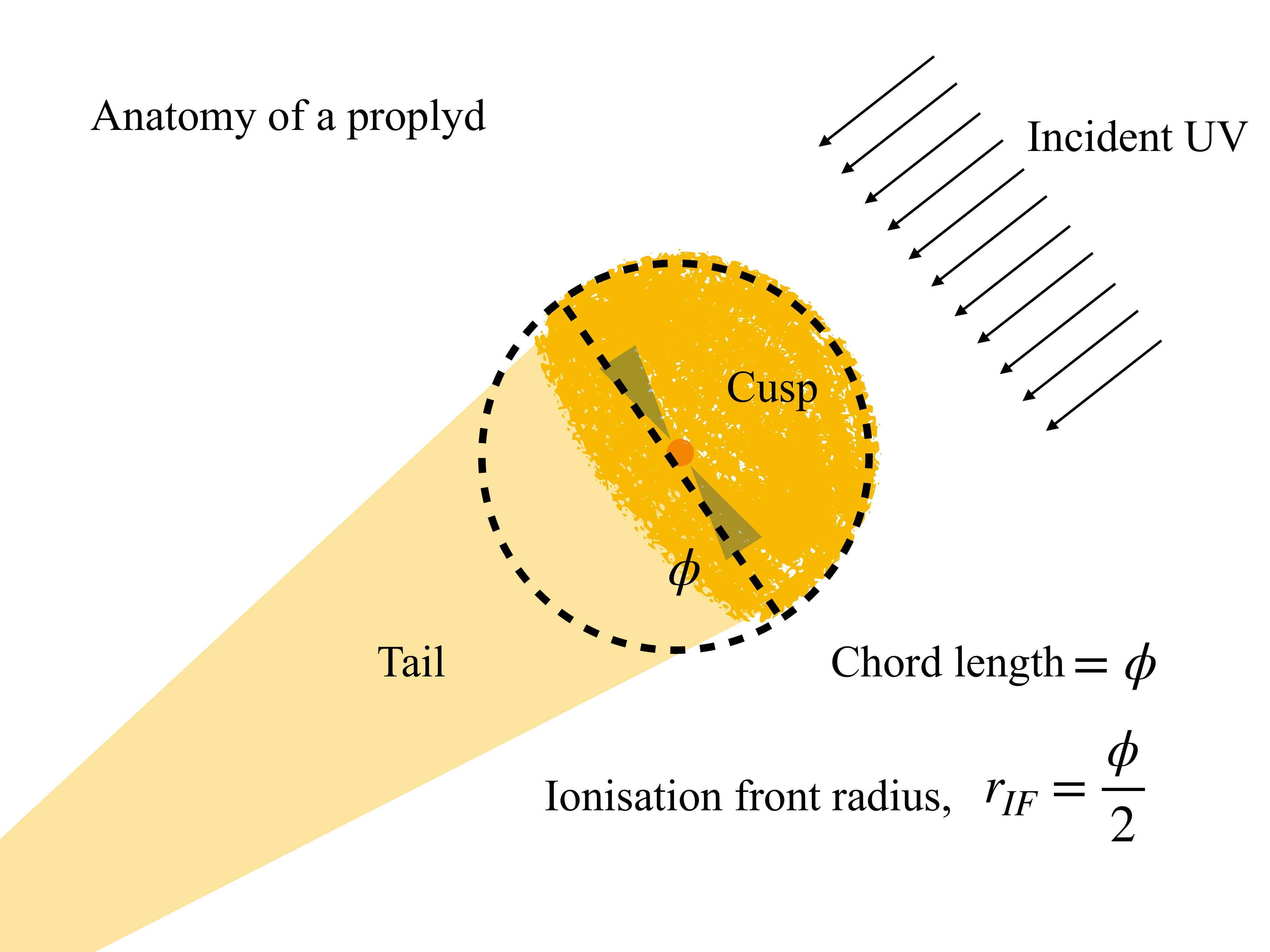}
    \caption{A schematic of a cometary proplyd. The disc is being irradiated by a UV field represented by the parallel arrows. This drives material from the disc into a bright cusp on the side facing the UV source and a tail on the side away from the source. When measuring the size of the proplyd (the ionisation front radius in our case) we use the radius of a circle that traces the cusp. }
    \label{fig:proplydschematic}
\end{figure}

We have growing statistical evidence for the impact of radiation environment on discs \citep[e.g.][]{2012A&A...539A.119F, 2014ApJ...784...82M, 2016ApJ...829...38M, 2016arXiv160501773G, 2017AJ....153..240A, 2018ApJ...860...77E}. However an ongoing issue is that we have very limited sample of individual discs with observed photoevaporative outflows. {The most intense study has been of photoevaporating} individual discs in the Orion Nebular Cluster (ONC) -- the famous ``proplyds'' with cometary morphologies consisting of a cusp and tail, as illustrated schematically in Figure \ref{fig:proplydschematic} \citep{1996AJ....111.1977M, 1998AJ....115..263O, 1999AJ....118.2350H,  2000AJ....119.2919B, 2002ApJ...566..315H}. {Other photoevaporating discs are in similarly strong, or stronger, UV environments including in Pismis 24 \citep[5 candidate proplyds in a UV environment even stronger than the ONC --][]{2012A&A...539A.119F} and Carina \citep[e.g.][]{2010MNRAS.405.1153S}}

Recently \cite{Kim16} discovered 7 proplyds in NGC 1977, a UV environment around 2 orders of magnitude weaker than the ONC. {This provided important first direct evidence for external photoevaporation operating in weaker UV environments.} However, more examples are required across a larger range of UV environments to truly understand the photoevaporation process in action. To this end we require both more proplyd detections and to understand evaporation signatures so that we can detect a wind when it is more subtle \citep[e.g. in CI and CO lines --][]{2020MNRAS.492.5030H}. 

The stellar cluster NGC 2024 was recently observed with ALMA by \cite{vanTerwisga20}, who found that there were two disc populations that are similar in age, but separated both spatially and in terms of disc fraction. The slightly younger ($\sim0.2-0.5$\,Myr) population, within and on the eastern side of a dense molecular cloud, has {an ALMA 1.3\,mm disc fraction of $45\pm7$\,per cent \citep[the rms noise of observations was 0.058\,mJy\,beam$^{-1}$ and lowest ``detection'' dust mass was $2.5\pm0.70$\,M$_\oplus$][]{vanTerwisga20}} and masses similar to  1-3\,Myr old star forming regions such as Lupus. The slightly older ($\sim1$\,Myr) population lies on the western side of the cloud and has {an ALMA mm} disc fraction of only $15\pm4$\,per cent. Being less shielded and containing the B0 star IRS 1 it was speculated that external photoevaporation may have been responsible for depleting the western disc population. However it is also possible that discs in the western region simply began with lower masses. {Note that the near infrared disc fraction over a much larger extent of the region ($6.7\times6.7\arcmin$ vs $2.9\times2.9\arcmin$) is higher, at around 85\,per cent \citep{2001AJ....121.1512H, 2001ApJ...553L.153H}}

\cite{2020arXiv200600019K} recently suggested that higher cosmic ray ionisation rates can enhance the impact of magnetic braking and lead to initially smaller (and hence possibly shorter lived) discs. They suggested that this might provide an alternative explanation for the different sub-mm detection fractions in NGC~2024. It is important to understand what has had such a large impact on these two disc populations, both to understand this region and the planet formation process more generally. Determining the impact of external photoevaporation in this type of environment is therefore key.  \\

Motivated by \cite{vanTerwisga20}, here we search archival Hubble Space Telescope (HST) data for photoevaporating discs in NGC 2024. We discover 4 proplyds and 4 further candidates which demonstrate that external photoevaporation is indeed operating throughout the region. 

\begin{figure*}
    \includegraphics[width=17.5cm]{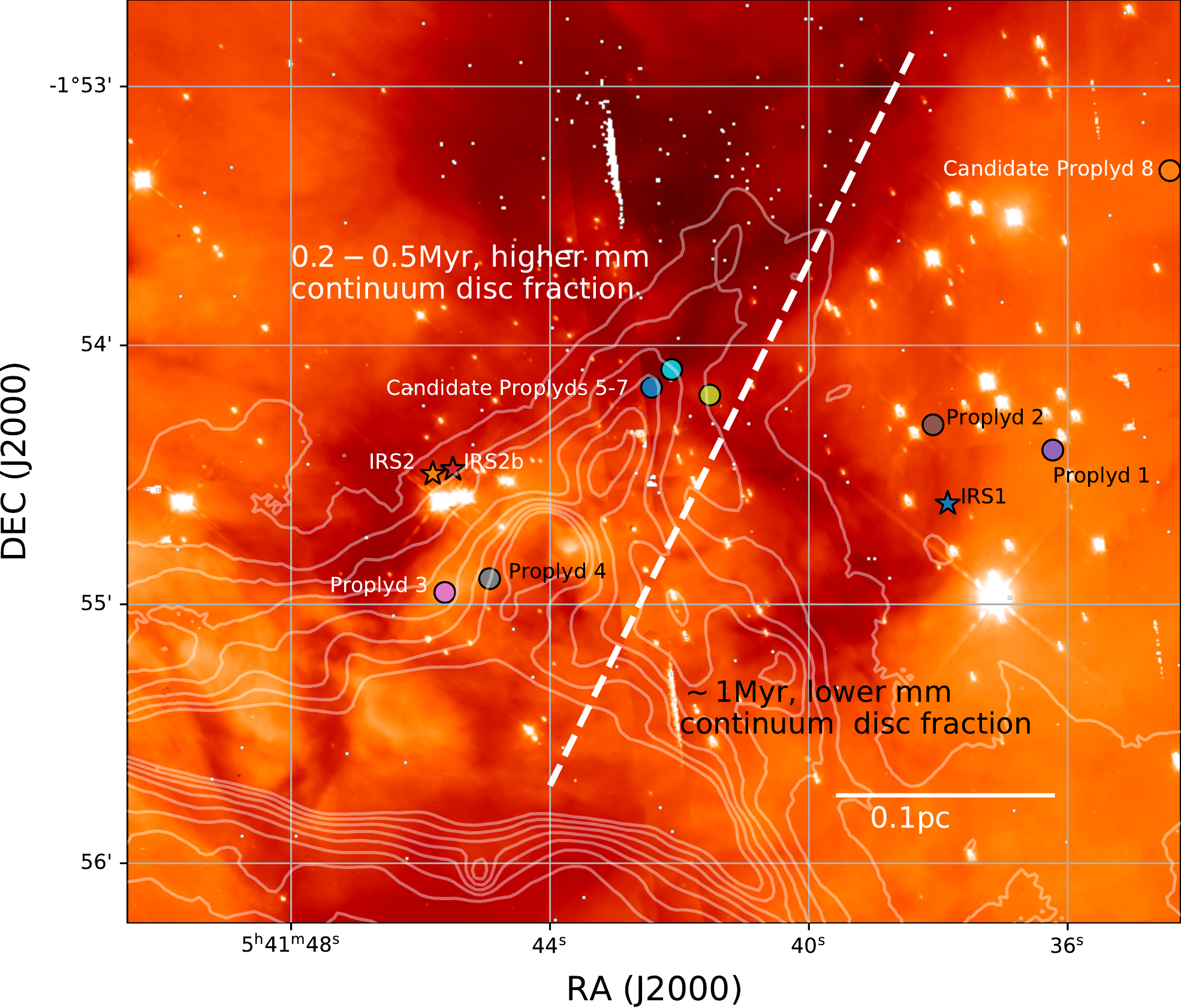}
    \caption{An overview of NGC 2024. The main image is a 1.6\,$\mu$m F160W HST mosaic taken from MAST \protect\citep{2016hst..prop14624A}. Overlaid are Herschel PACS blue 70um contours, from 1-6 Jy/pixel \protect\citep{2013ApJ...767...36S}. The 60\arcsec\ grid is 0.12\,pc at a distance of 414\,pc \protect\citep{2007A&A...474..515M, 2018AJ....156...58B}. Star symbols denote the main UV sources (IRS1 and IRS2). Circles denote the three proplyds introduced in this paper. The diagonal dashed line separates the two populations of marginally different age, but substantially different disc fraction \protect\citep{vanTerwisga20}.}
    \label{fig:NGC2024}
\end{figure*}

\section{Overview of UV sources and structure of NGC~2024}
\label{sec:NGC2024}
A mosaic overview of NGC 2024 in the HST F160W filter ($\sim1.6\mu$m) constructed with data from the MAST\footnote{\url{https://archive.stsci.edu/}} science archive (PI: Arce) is given in Figure \ref{fig:NGC2024}. Overlaid are Herschel PACS blue 70$\mu$m contours from 1--6\,Jy\,pixel$^{-1}$ using data from the ESA Herschel Science Archive\footnote{\url{http://archives.esac.esa.int/hsa/whsa/}} (PI: Megeath). The diagonal line at the western edge of the cloud separates the younger ($\sim0.2-0.5$\,Myr) population from the 1 Myr old population that exhibit relatively depleted mm-fluxes \citep{2000AJ....120.1396H, 2014ApJ...787..109G, vanTerwisga20}. The star symbols denote the positions of the main UV sources: IRS1 (B0.5V star), IRS2b (08V star) and IRS 2 (B star), which are summarised in Table \ref{tab:UVsources}. The circles denote the positions of proplyds introduced in this paper (discussion of these is left for sections \ref{sec:proplyds} and \ref{sec:discussion}. 

NGC~2024 is the youngest cluster in Orion and richest in star formation in Orion B \citep{1996PhDT.........3M, 2006ApJ...646.1215L}. There are two main UV sources.  IRS 1 \citep[a B0.5V star --][]{1968PASP...80...20G} is not embedded and resides in the $\sim1$\,Myr region which \cite{vanTerwisga20} conclude has a depleted mm disc population. Given our interest in proplyds it is worth noting that IRS 1 is similar to 42 Ori in NGC 1977, where 8 proplyds have now been discovered \citep{2012ApJ...756..137B, Kim16}. 

IRS 2b \citep[an O8V star][]{2003A&A...404..249B} is in the eastern ($\sim0.2-0.5$\,Myr) population \citep{vanTerwisga20}. IRS 2 is an embedded B-type star in close projected proximity to IRS 2b and is expected to be of secondary importance in terms of its contribution to the UV flux in the region. 

For some time only IRS 1 and 2 were known in NGC 2024, giving rise to the problem that neither have hard enough a spectrum to explain the degree of hydrogen and helium ionisation \citep{2003A&A...404..249B}. The discovery of the more massive IRS 2b appears to resolve this \citep[e.g.][]{1981ApJ...249..622T, 1982A&AS...48..345K, 1989ApJ...342..883B, 2003A&A...404..249B, 2007PASJ...59..487K, 2012ApJ...756L...6B}, though it is still not ruled out that there are perhaps other UV sources on the far side of the dense cloud. \cite{1982A&AS...48..345K} and \cite{2003A&A...404..249B} respectively suggest $\sim10^{48}$ and $7.3\pm1.2\times10^{47}$ ionising photons per second are being absorbed by the gas to give the observed radio continuum signal. Note that these values are a lower limit since dust absorption of ionising photons, or ionising photons escaping from the region, will not ionise the gas and contribute to the radio continuum. IRS 1 cannot explain all of the ionising flux expected from the radio continuum, but the exact value of the relative contributions from IRS 1/ IRS 2b is unknown.

In many star forming regions a single ionising source is dominant. However, NGC 2024 has a complicated structure with a lot of gas and dust in the region, so it isn't obvious that IRS 2b necessarily dominates the UV of the entire region. 

The cusp of proplyds point towards the UV source responsible for evaporating them (see Figure \ref{fig:proplydschematic}), so the detection of proplyds in NGC 2024 would also shed light on where the different UV sources dominate over one another. Furthermore, if proplyds were discovered pointing towards none of the known UV sources, it would give clues to the location of additional massive star(s).

\section{Theoretical potential for proplyds}
For proplyds to be visible and resolvable requires a combination of sufficient EUV irradiation and a high-enough mass-loss rate. The radius $R_\mathrm{IF}$ of the ionisation front can be approximated by assuming a balance between EUV flux and recombinations in the flow \citep[see][]{Johnstone98}:
\begin{multline}
\label{eq:RIF}
    R_\mathrm{IF} \approx 1200 \, \left( \frac{\dot{M}_\mathrm{wind}}{10^{-8} \, M_\odot \, \mathrm{yr}^{-1}}\right)^{2/3} \\ 
    \times \left( \frac{\dot{N}_\mathrm{Ly}}{10^{45} \, \mathrm{s}^{-1}}\right)^{-1/3} \left( \frac{d}{1\,\mathrm{pc}} \right)^{2/3} \, \mathrm{au}
\end{multline} 
where $\dot{N}_\mathrm{Ly}$ is the EUV photon count from the (external) ionising source, $d$ is the distance to that source, and $\dot{M}_\mathrm{wind}$ is the wind driven mass-loss rate from the disc.

We estimate the radiation field at any given point in NGC 2024 as a function of the distance (in practice, the projected separation) from the two known main UV sources: IRS 1 and IRS 2b (see section \ref{sec:NGC2024}/Table \ref{tab:UVsources}).

\begin{table}
    \centering
     \caption{The main known UV sources in NGC 2024.}
    \begin{tabular}{c|c|c|c|c|c}
    \hline
      UV   & RA & DEC & Spectral &   \\
      Source & & & class &   \\
      \hline
      IRS 1   &	05 41 37.8540 & -01 54 36.6256 & B0.5V  \\
      IRS 2 & 05 41 45.81 & -01 54 29.8 & B \\
      IRS 2b &	05 41 45.50  & -01 54 28.7 &  O8V   \\
    \hline
    \end{tabular}
    %\caption{TJH Coordinates of some points of interest}
    \label{tab:UVsources}
\end{table}

The intrinsic hydrogen ionising photon count from these sources, following \cite{2003ApJ...599.1333S}, is $\log_{10}(\dot{N}_\mathrm{Ly, IRS1}) = 47.71$ and $\log_{10}(\dot{N}_\mathrm{Ly, IRS2b}) = 48.75$ for IRS 1 and IRS 2b respectively. This does give a higher input ionising flux than inferred from the radio continuum \citep[e.g.][]{2003A&A...404..249B}, but as discussed in section \ref{sec:NGC2024} the radio continuum does only provide a lower limit. In the discussion section \ref{sec:Mdots} we will also consider values a factor 5 lower for the ionising photon emission rate from IRS1/IRS2b. 

As a function of distance $d$ (projected separation) we hence have an ionising photon flux of 
\begin{equation}
   \frac{\dot{N}_\mathrm{Ly, IRS1}}{4\pi d^2} = 4.3\times10^{11}\left(\frac{d}{0.1\,\textrm{pc}}\right)^{-2}\,\textrm{s}^{-1}\textrm{cm}^{-2}. 
      \label{equn:Nly1}
\end{equation}
and
\begin{equation}
   \frac{\dot{N}_\mathrm{Ly, IRS2b}}{4\pi d^2} = 4.7\times10^{12}\left(\frac{d}{0.1\,\textrm{pc}}\right)^{-2}\,\textrm{s}^{-1}\textrm{cm}^{-2}. 
      \label{equn:Nly2b}   
\end{equation}
for IRS1 and IRS2b respectively.

The ionising flux is important for estimating mass loss rates as per equation \ref{eq:RIF}. It is also useful to estimate the magnitude of the FUV radiation, which we do to first order by integrating over a blackbody spectrum, assuming stellar temperature and luminosity from \cite{2003ApJ...599.1333S}.  This gives $L_\mathrm{FUV, IRS 1} \sim 1.4 \times10^{38}$~erg~s$^{-1}$ and $L_\mathrm{FUV, IRS 2b} \sim 3.4 \times10^{38}$~erg~s$^{-1}$ for IRS1 and IRS 2b respectively. IRS1 is in a relatively open part of the region, so there is comparatively little extinction to the stellar neighbours in the cluster. Extinction may play a large role in the vicinity of IRS2b though, with many of the nearby stars being embedded. Assuming the no-extinction case provides an upper limit, the FUV radiation field strength as a function of distance from the UV sources is
\begin{equation}
   \mathrm{FUV}_\mathrm{IRS1} \leq 7.34\times10^4\left(\frac{d}{0.1 \, \textrm{pc}}\right)^{-2}\,G_0
   \label{equn:FUV1}
\end{equation}
and  
\begin{equation}
   \mathrm{FUV}_\mathrm{IRS2b} < 1.8\times10^5\left(\frac{d}{0.1 \, \textrm{pc}}\right)^{-2}\,G_0
     \label{equn:FUV2b} 
\end{equation}
for IRS1 and IRS2b respectively. where $G_0$ is the Habing unit of the radiation field\footnote{Note that an alternative measure of the UV field is the Draine $\chi$, which is 1.71 times larger than the Habing unit. Hence $1\, G_0$ is $\sim0.585$\,$\chi$}.

Given that the region studied by \cite{vanTerwisga20} and illustrated in Figure \ref{fig:NGC2024} is around half a parsec in size, these ionising and FUV flux estimates are high enough that we  theoretically expected them to drive significant photoevaproative winds from discs in the region, as long as they are not deeply embedded \citep[e.g.][]{Johnstone98, 2018MNRAS.481..452H}.

\section{Archival HST Observations}
We searched for historical Hubble Space Telescope (HST) observations towards NGC 2024 in the MAST\footnote{\url{https://archive.stsci.edu/}} science archive and Hubble Legacy Archive\footnote{\url{https://hla.stsci.edu/}} (HLA).

\subsection{WFC2 H\,$\alpha$}
We studied Wide Field Camera 2 H\,$\alpha$ images from the HLA. For these all we required was the morphological identification of proplyds, their coordinates and measurements of the ionisation front size. We therefore used the data directly as it appears in the science archive, with no further manipulation. We use a single field that is image 1 from Proposal ID 5983, F656N, PI: Stapelfeldt 1995. {The pixel size is 0.1\arcsec which corresponds to 41\,au at a distance of 414\,pc \citep[][]{2007A&A...474..515M, 2018AJ....156...58B}}.

\subsection{NICMOS Paschen\,$\alpha$}
The main dataset that we use was originally obtained in 2003 by Stapelfeldt \& Hines (Proposal ID: 9424, PI Stapelfeldt). These used camera 2 of NICMOS in filters centered on Paschen\,$\alpha$ (F187N) and the nearby (in wavelength) continuum (F190N) to image fields centered on 16 compact VLA sources in the region. Subtracting the F190N image from the F187N therefore leaves continuum subtracted Paschen\,$\alpha$ emission. {The pixel size is 0.075\arcsec which corresponds to 31\,au at a distance of 414\,pc}.

A 3-position spiral dither pattern was executed for each
object, with 5\arcsec\ offsets to mitigate against bad pixels, residual cosmic rays and other
imaging artifacts within 10\arcsec\ of the centre of the field of view. Three positions (the usual
recommended pattern uses four positions) were used to maximize the serendipitous field of view and
yet still fit within the allocated observing time. After processing through the STScI pipeline, the three dithered images were shifted into
common world coordinates and median combined. The resulting mosaic images were used
to construct Paschen\,$\alpha$ line images by subtracting the F190N image scaled by the
appropriate bandpass (the stellar photosphere is assumed to be flat across the F187N and
F190N bandpasses).

%{TJH I dont really know to what level of detail we need to go into here. H alpha proplyds pretty much just using the HLA data as is. Paschen alpha we did the subtraction + there are the details about artefacts. Input very welcome in this section. }

%\SK{For H$\alpha$ I think we can just say that we have downloaded the image from HLA and give the PI name and proposal ID (PID 5983, PI Stapelfeldt, did we also use H$\alpha$ from PID 9424?) and observed dates.  For Paschen, Dean's draft had details and I think that would suffice for this paper (since we did not work on enhancing/improving the data for this paper).  I will get back to this...}

\begin{figure*}
    \hspace{-1cm}
    \includegraphics[width=1.6\columnwidth]{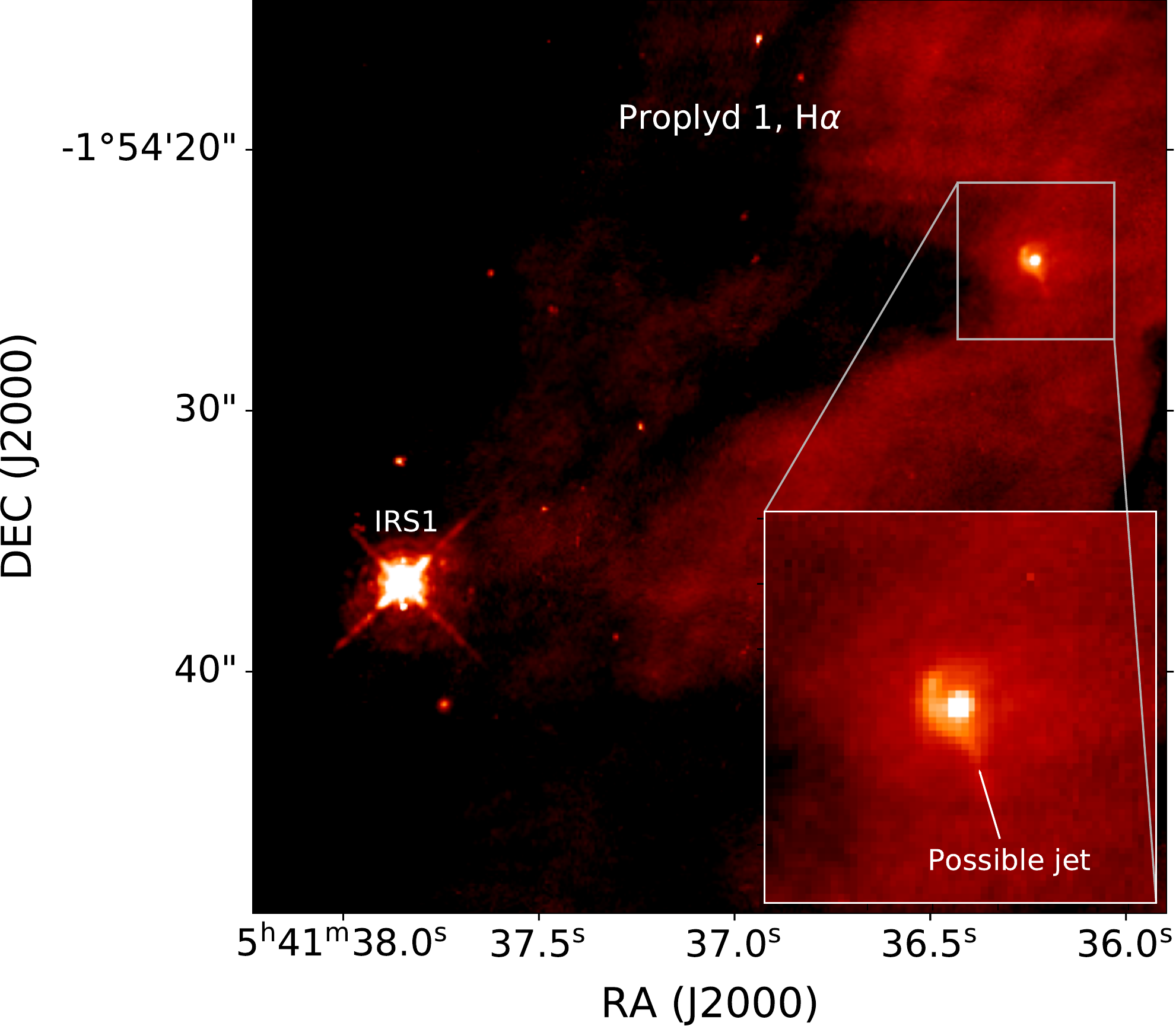}
    \caption{Proplyd 1, discovered in an archival HST WFC2 F656N (H\,$\alpha$) image. The star responsible for evaporating it is IRS 1, a B0.5V star situated only 0.055\,pc away. Included in the inset box is a zoom in on the proplyd itself. The material almost perpendicular to the line connecting IRS1 and the bright proplyd cusp is likely a stellar jet. }
    \label{fig:Proplyd1}
\end{figure*}

\begin{table*}
    \centering
     \caption{Properties of proplyds in NGC 2024. Columns are the proplyd ID, coordinates, the UV source responsible for the photoevaporation, projected separation from that source (assuming a distance of 414\,pc), ionisation front radius, estimated mass loss rate, any counterpart and the corresponding dust mass from the \protect\cite{vanTerwisga20} dataset. Note that the Paschen $\alpha$ pixel size is 31\,au at 414\,pc.  }
    \begin{tabular}{c|c|c|c|c|c|c|c|c|c|c}
    \hline
          \hline
      Object   & RA & DEC &UV  & Projected    & $R_{if}$    & $\dot{M}$ &Van Terwisga et al. & $M_d$  \\
      & & & source & Separation (pc)   &  (au)  &  ($10^{-7}M_\odot$\,yr$^{-1}$)  & 2020 counterpart  & ($M_\oplus$) \\
      \hline
            \hline
      Proplyd 1   & 05 41 36.23  & -01 54 24.26 &IRS1 & 0.055 & $144\pm62$ & $1.7^{+1.2}_{-0.9}$ &IRC215 & $67.2 \pm 0.77$  \\   
      (VLA 1) & & & & (0.28 from IRS 2b) \\
      \hline      
      Proplyd 2    & 05 41 38.08  & -01 54 18.41 &IRS1 & 0.037 &$206\pm 31$ & $4.3\pm1.0$  &IRC116 & $14.6 \pm 0.72$  \\
      (VLA 4) & & & & (0.22 from IRS2b) & &  \\
      \hline
      Proplyd 3 & 05 41 45.62 & -01 54 57.26 &IRS2b & 0.057 & $93\pm31$ & $2.8^{+1.5}_{-1.3}$  &IRC057 & $13.8 \pm 2.56$ \\
      (VLA 20b) & & & & (0.23 from IRS1) & &  \\
            \hline
      Proplyd 4 & 05 41 44.93 &  -01 54 54.06 & IRS2b & 0.054 & $87\pm37$  & $2.7^{+1.9}_{-1.5}$ &IRC59 & $26.8 \pm 0.87$ \\
      (VLA 20a) & & & & (0.22 from IRS1) & \\
            \hline
      Candidate  & 05 41 41.53  & -01 54 11.5 & IRS2b & 0.124 & $171^{+78}_{-31}$ & $3.2^{+5.7}_{-2.4}$& IRC 123 & $40.4 \pm 0.71$  \\ 
      Proplyd 5 & & & & (0.121 from IRS1)  \\
            \hline
      Candidate  & 05 41 42.12  & -01 54 05.58  & IRS2b& 0.11& $77\pm31$ & $1.1^{+0.7}_{-0.6}$ & -- & --   \\ 
      Proplyd 6 & & & & (0.14 from IRS1)  \\
            \hline
      Candidate  & 05 41 42.43   &  -01 54 09.72 & IRS2b& 0.10 & $109\pm31$ &$2.0^{+0.9}_{-0.8}$ & IRC 124 & $42.5 \pm 1.79$    \\ 
      Proplyd 7 & & & & (0.15 from IRS1) \\
            \hline
      Candidate   & 05 41 34.42   & -01 53 19.49 &IRS1 & 0.19  & $<60$ & $<0.1$ & -- & --   \\   
      Proplyd 8 & & & & (0.36 from IRS 2b) & & &  \\
            \hline
    \hline
    \end{tabular}
    %\caption{TJH Coordinates of some points of interest}
    \label{tab:ProplydParams}
\end{table*}

\section{Proplyds in NGC 2024}
\label{sec:proplyds}

\subsection{Overview}
We have firmly identified 4 proplyds and have 4 other candidate proplyds in NGC 2024. Two of these were found in the F656N filter (H\,$\alpha$) and the other 6 in F187N-F190N (Paschen\,$\alpha$). Three proplyds are in the western region \citep[the $\sim1\,$Myr population with only $\sim15$\,per cent of discs detected in the mm-continuum by][]{vanTerwisga20} and pointing towards the B0.5V star IRS 1. The other 5 are in the $\sim0.2-0.5\,$Myr higher {mm} disc fraction region, pointing towards the O8 star IRS2b. These will all be discussed in detail below, but for quick reference, image galleries of the proplyds are given in Figures \ref{fig:Proplyd1}--\ref{fig:Proplyds5To8} and a summary of their main parameters is given in Table \ref{tab:ProplydParams}. The locations of the 8 propylds and the main UV sources are also provided in the overview of the region in Figure \ref{fig:NGC2024}. 

We next introduce the basic properties (morphology, location, environment) of each proplyd in turn and in the discussion turn our attention to the region as a whole. 

The ionisation front radii were estimated using the radius of a circle manually drawn over the cusp of the proplyd (see Figure 1). Uncertainties are based on circles that trace the inner and outer edge of the cusp.

\subsection{Proplyd 1}

Proplyd 1, discovered in F656N H\,$\alpha$ \citep[and preliminarily introduced in the proceedings of ][]{1997ASPC..119..131S} is shown in Figure \ref{fig:Proplyd1}. 
It is at a projected separation \citep[{assuming a distance of 414\,pc,}][]{2007A&A...474..515M, 2018AJ....156...58B} of 0.055\,pc from the B0.5V star IRS 1, towards which its bright cusp is directed. The morphology of Proplyd 1 is not simply cometary, with additional H\,$\alpha$ emission extending to the south west. We interpret this as being from a jet. 

From equations \ref{equn:Nly1} and \ref{equn:FUV1} the UV and ionising flux is $2.4\times10^6$\,$G_0$ and $1.4\times10^{12}$\,s$^{-1}$ respectively. From an approximate manual measurement of the radial extent of the cusp we assume an ionisation front radius of $144\pm62$\,au. The expected mass loss rate from equation \ref{eq:RIF} for this proplyd is hence $1.7^{+1.2}_{-0.9}\times10^{-7}$\,M$_\odot$\,yr$^{-1}$. 
Proplyd 1 also has a counterpart in \cite{vanTerwisga20}, where a dust mass of $67.2\pm0.77$\,M$_\oplus$ is inferred. 

\begin{figure}
    \hspace{-1cm}
    \includegraphics[width=1.\columnwidth]{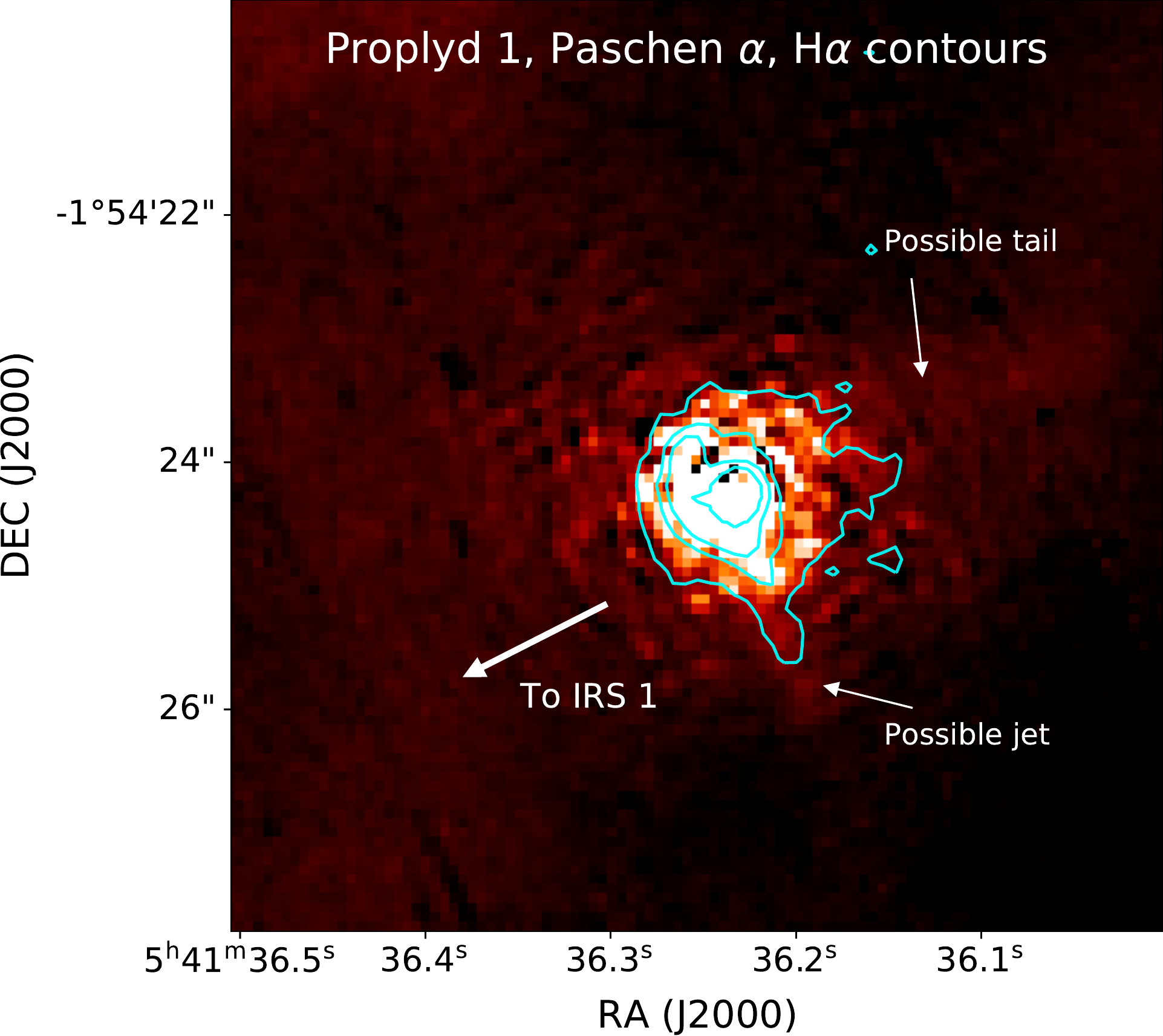}
    \caption{{The Paschen\,$\alpha$ data for proplyd 1 with the H\,$\alpha$ contours overlaid (see Figure \ref{fig:Proplyd1}). In this case the Paschen\,$\alpha$ emission from the star coupled with a slightly different PSF for F187N and F190N leads to artifacts, but extended emission from the jet and possibly a longer component of the tail are still visible.  }}
    \label{fig:Proplyd1wPa}
\end{figure}

{There is also archival F187N-F190N (Paschen\,$\alpha$) data for proplyd 1, which is shown in Figure \ref{fig:Proplyd1wPa} with H\,$\alpha$ contours overlaid. Since the point spread function of the F187N and F190N images are slightly different sizes, unresolved Paschen\,$\alpha$ emission associated with the star has led to artifacts which makes the cusp and jet difficult to identify. However, larger scale emission from the jet and possibly also a more extended component of the tail are discernible. }

\subsection{Proplyd 2}
Proplyd 2 was discovered in F187N-F190N (Paschen\,$\alpha$) emission. The F187N, F190N (continuum) and F187N-F190N images for Proplyd 2 are shown in the upper row of panels in Figure \ref{fig:Proplyds2to4}. Proplyd 2 is 0.037\,pc from IRS 1 and an arrow is included on each panel pointing towards it. Like Proplyd 1, there are bipolar streaks of emission in addition to the typical proplyd cusp which as in the case of proplyd 1 we again interpret as being due to a jet. 

From equations \ref{equn:Nly1} and \ref{equn:FUV1}  the UV and ionising flux is $5.4\times10^6$\,$G_0$ and $3.1\times10^{12}$\,s$^{-1}$ respectively. From an approximate manual measurement of the radial extent of the cusp we assume an ionisation front radius of $206\pm31$\,au. The expected mass loss rate from equation \ref{eq:RIF} for this proplyd is hence $4.3\pm1.0\times10^{-7}$\,M$_\odot$\,yr$^{-1}$. 

Proplyd 2 also has a counterpart in \cite{vanTerwisga20}, where a dust mass of $14.6\pm0.72$\,M$_\oplus$ is inferred.

\subsection{Proplyd 3}
Proplyd 3 was discovered in F187N-F190N (Paschen\,$\alpha$) emission. The F187N, F190N (continuum) and F187N-F190N images for Proplyd 3 are shown in the middle row of panels in Figure \ref{fig:Proplyds2to4}. Proplyd 3 is 0.057\,pc from the 08V star IRS 2b, towards which it points. An arrow is included on each panel in the direction of IRS 2b. There are no additional ``jet'' features for Proplyd 3. It beautifully exhibits the classical cometary morphology of many ONC proplyds \citep[e.g.][]{1998AJ....115..263O}. 

From equations \ref{equn:Nly1} and \ref{equn:FUV1} the UV and ionising flux is $5.5\times10^6$\,$G_0$ and $1.4\times10^{13}$\,s$^{-1}$ respectively. From an approximate manual measurement of the radial extent of the cusp we assume an ionisation front radius of $93\pm31$\,au. The expected mass loss rate from equation \ref{eq:RIF} for this proplyd is hence $2.8^{+1.5}_{-1.3}\times10^{-7}$\,M$_\odot$\,yr$^{-1}$. 

Proplyd 3 also has a counterpart in \cite{vanTerwisga20}, where a dust mass of $13.8\pm2.56$\,M$_\oplus$ is inferred.

\subsection{Proplyd 4 (and possibly 4b)}
\begin{figure*}
    \hspace{-1cm}
    \includegraphics[width=5.86cm]{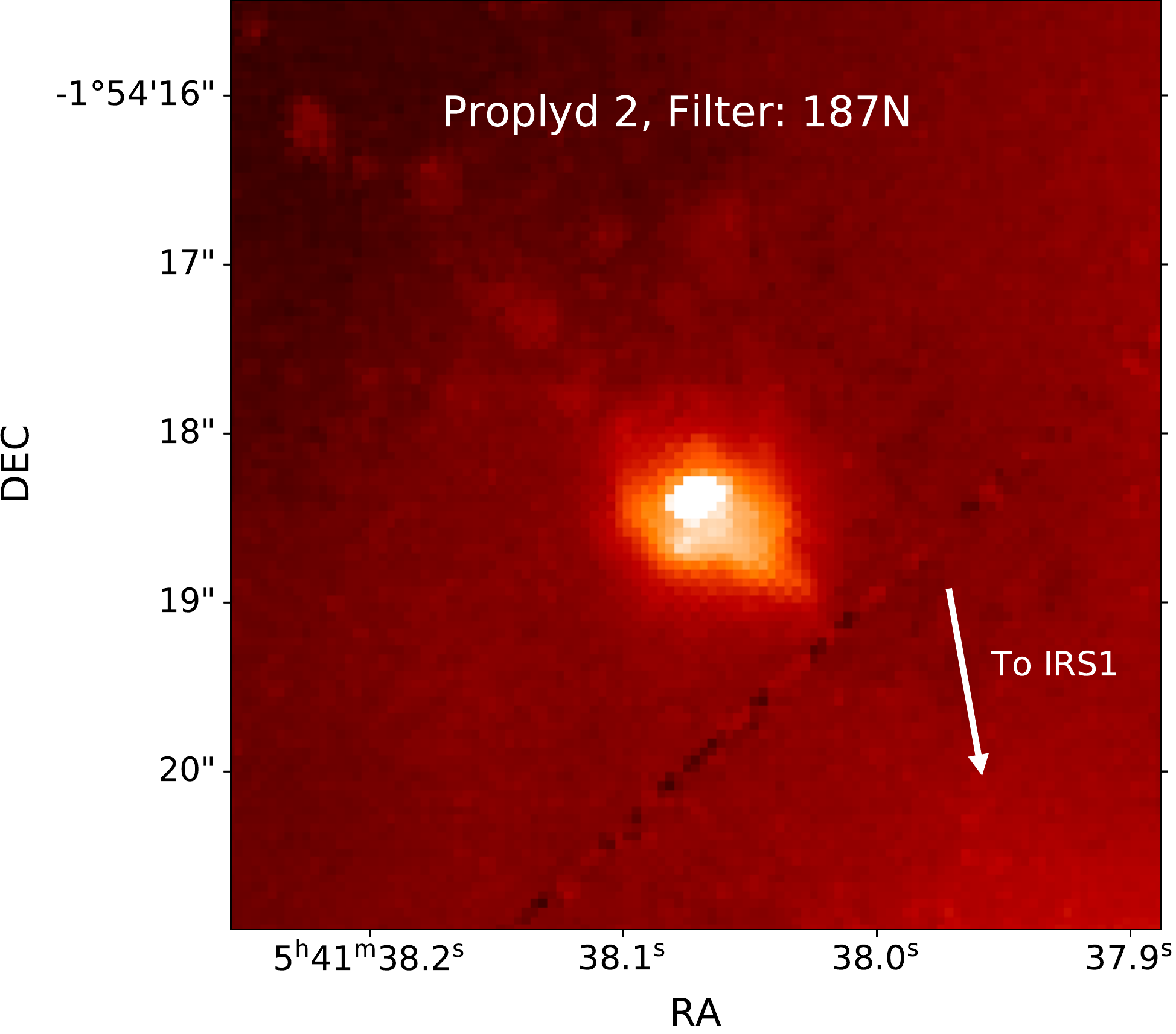}
    \includegraphics[width=5.86cm]{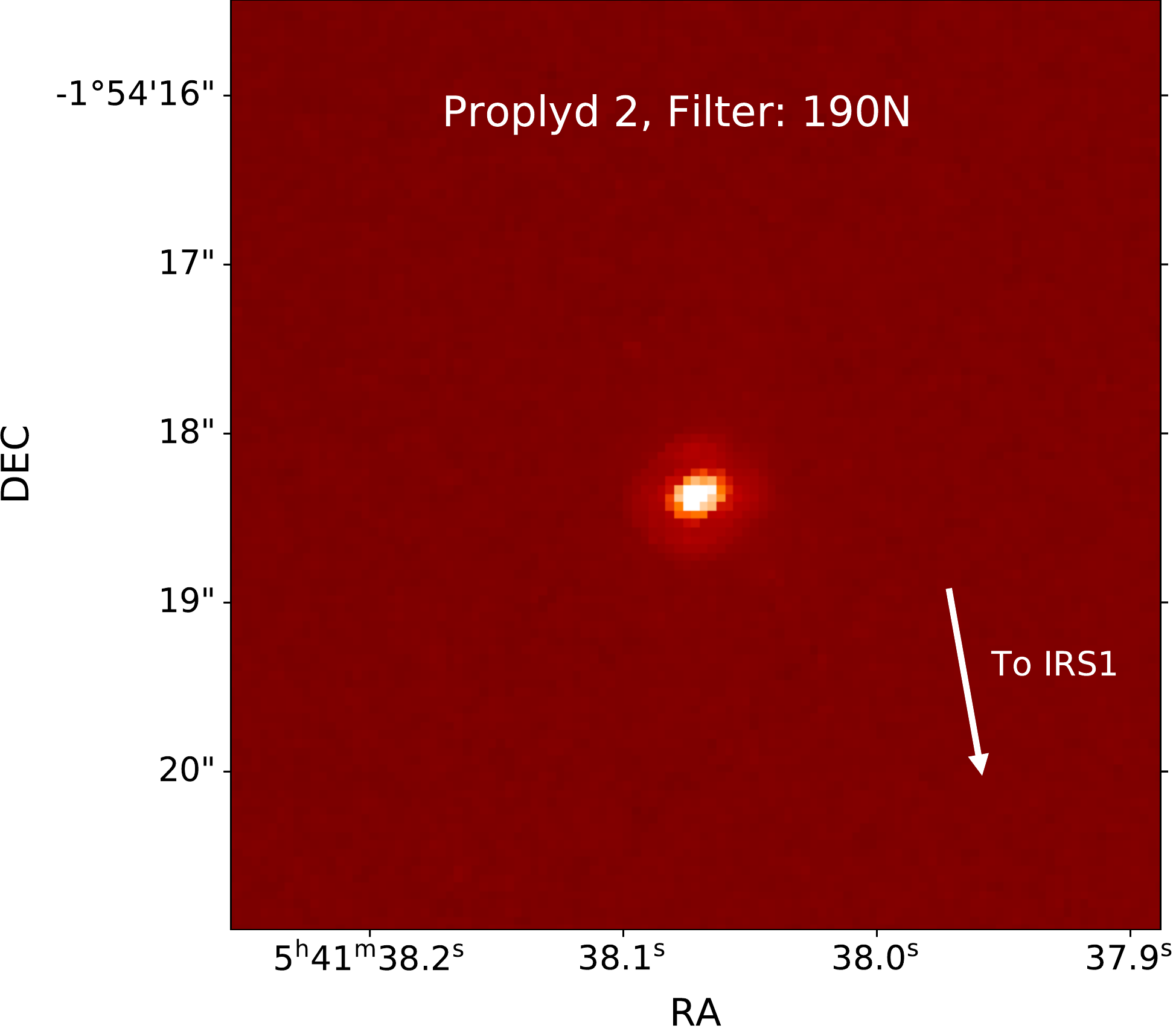}
    \includegraphics[width=5.86cm]{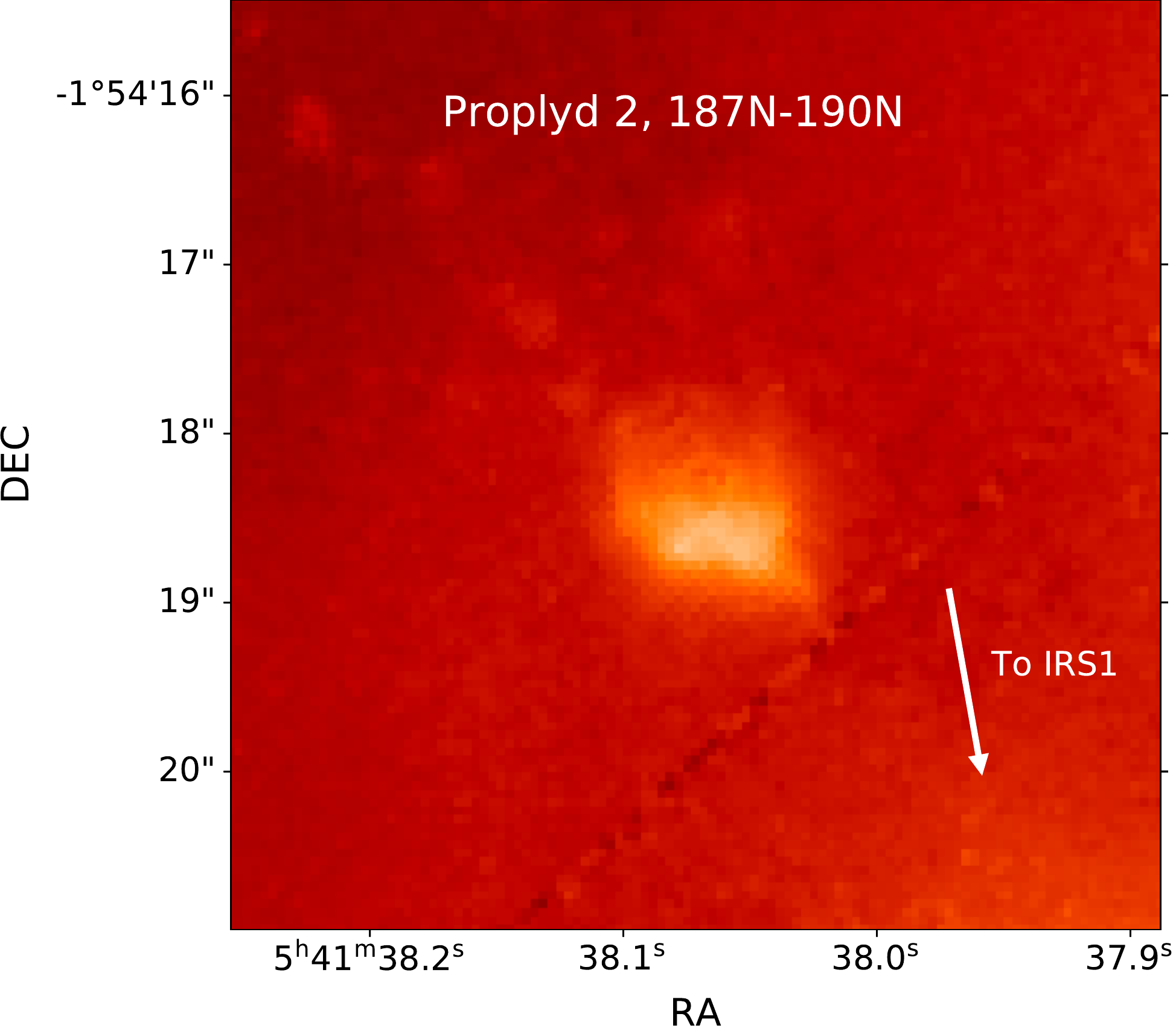}

    \hspace{-1cm}
    \includegraphics[width=5.86cm]{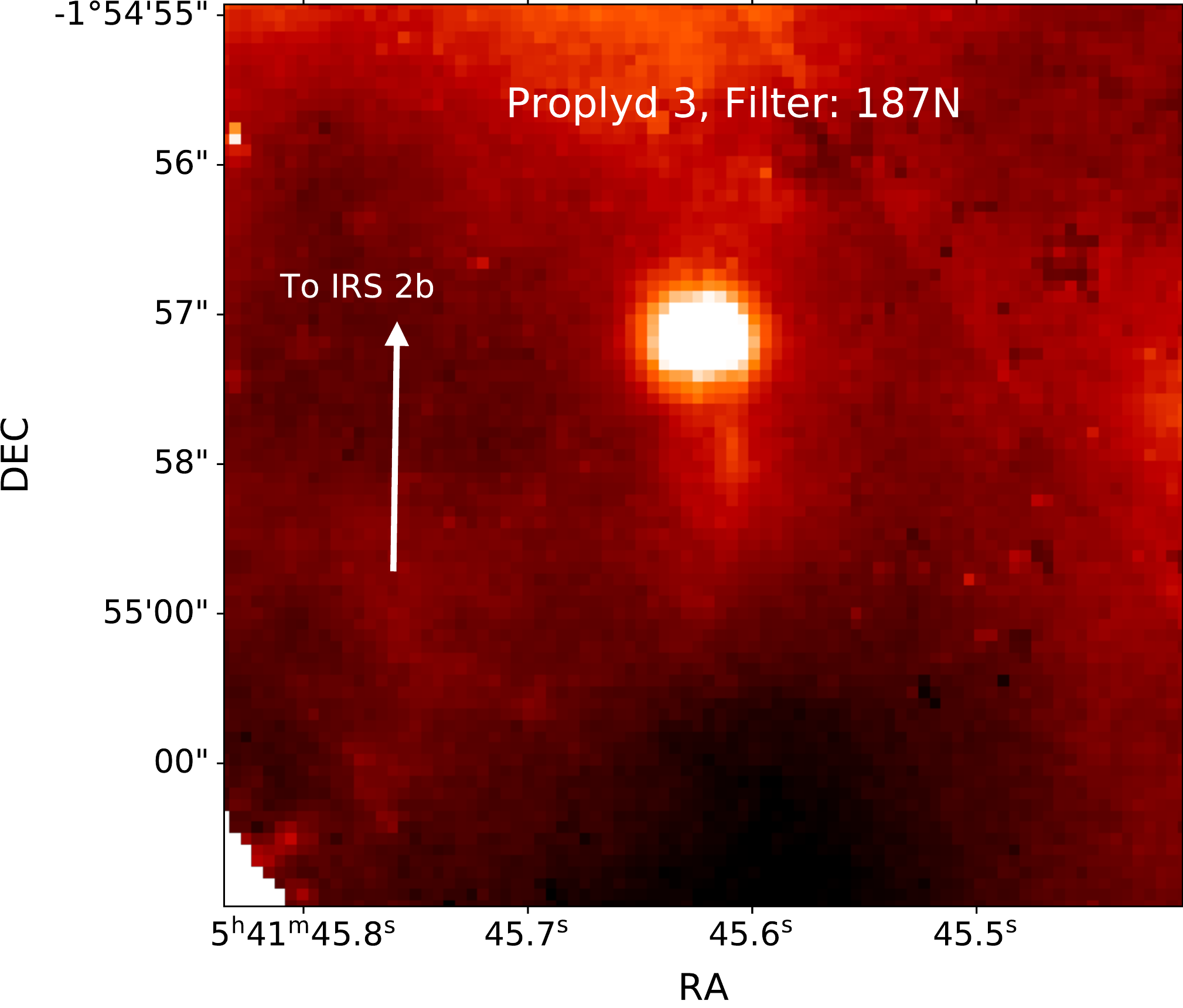}
    \includegraphics[width=5.86cm]{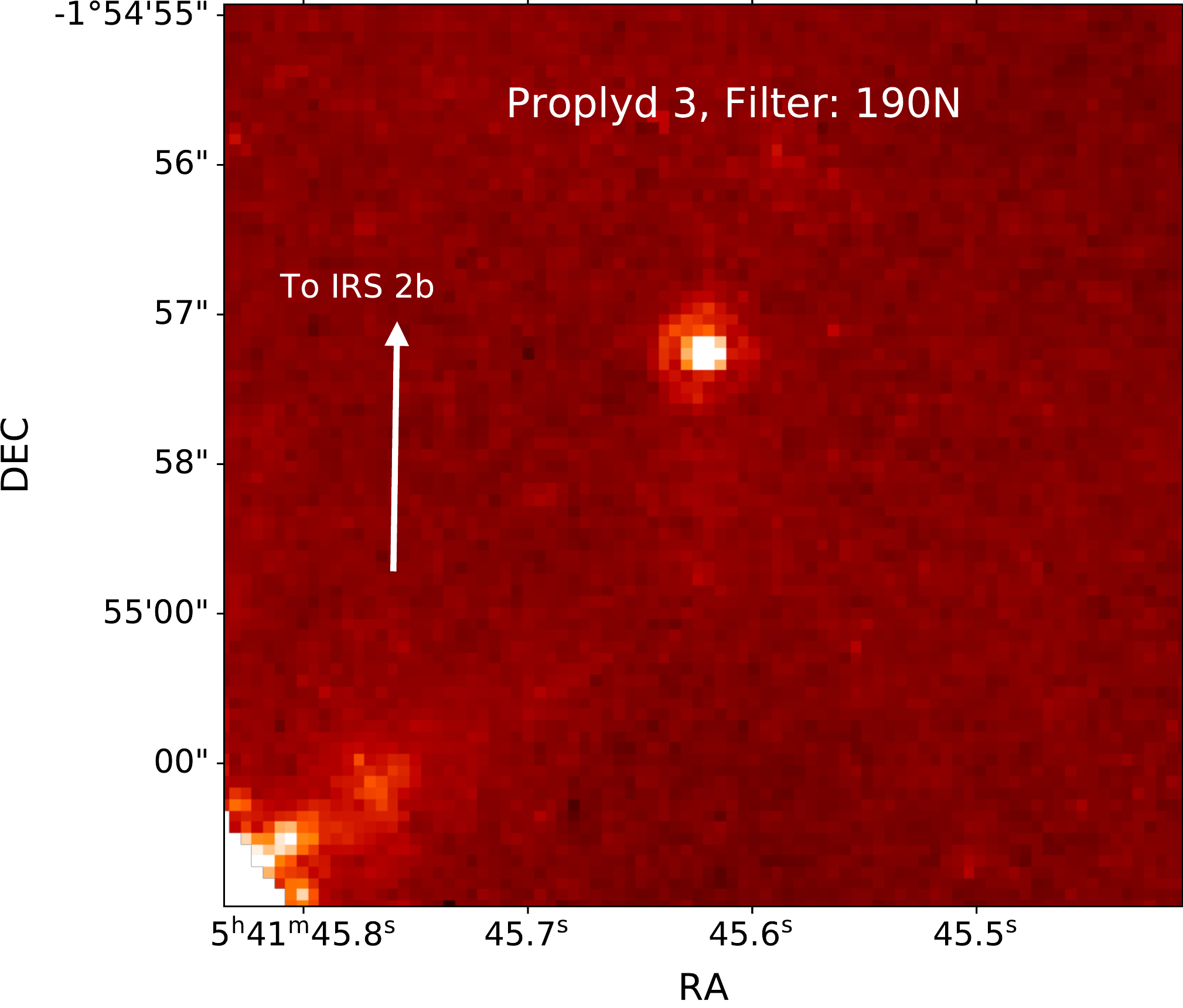}
    \includegraphics[width=5.86cm]{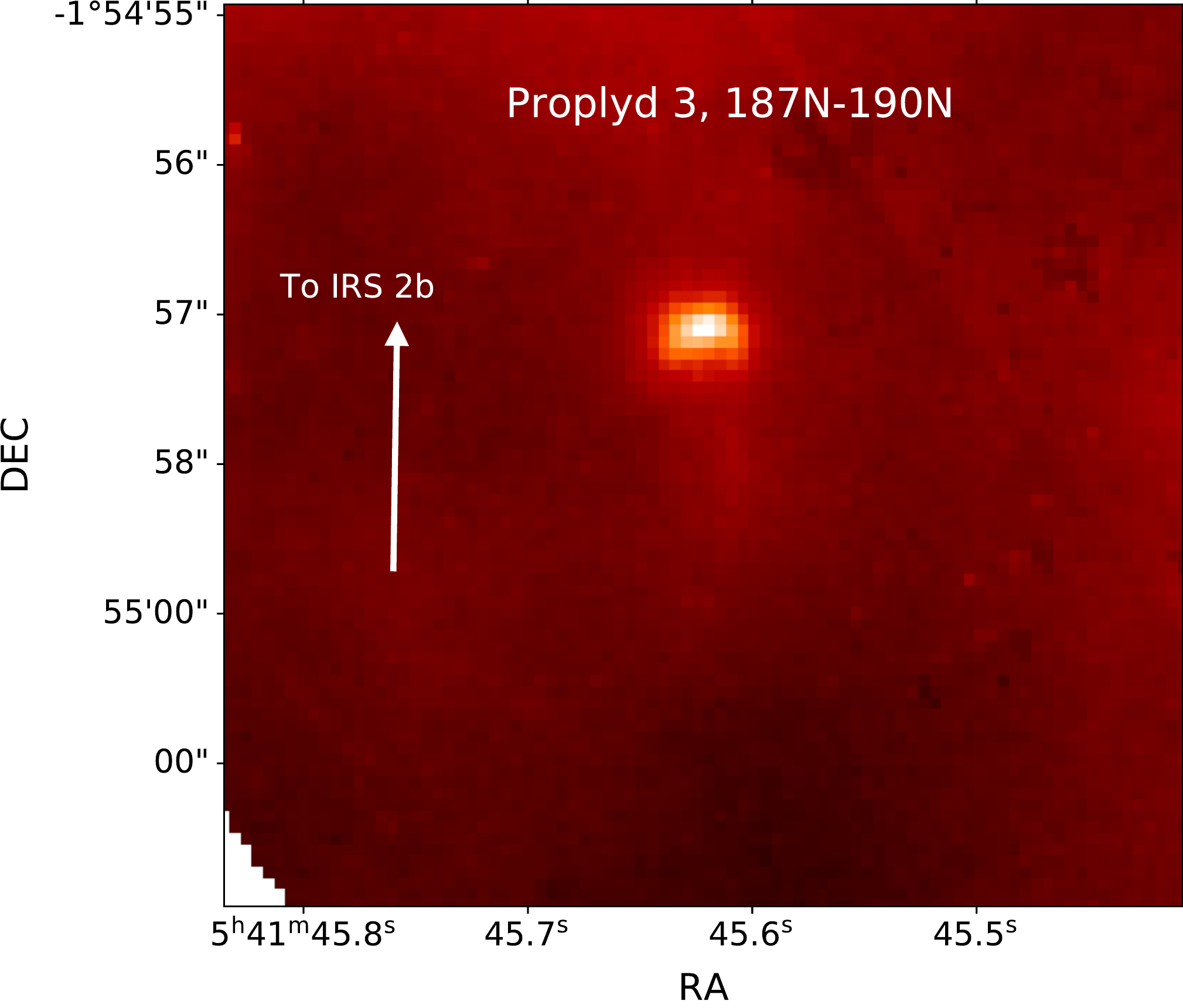}    
    
    \hspace{-1cm}
    \includegraphics[width=5.86cm]{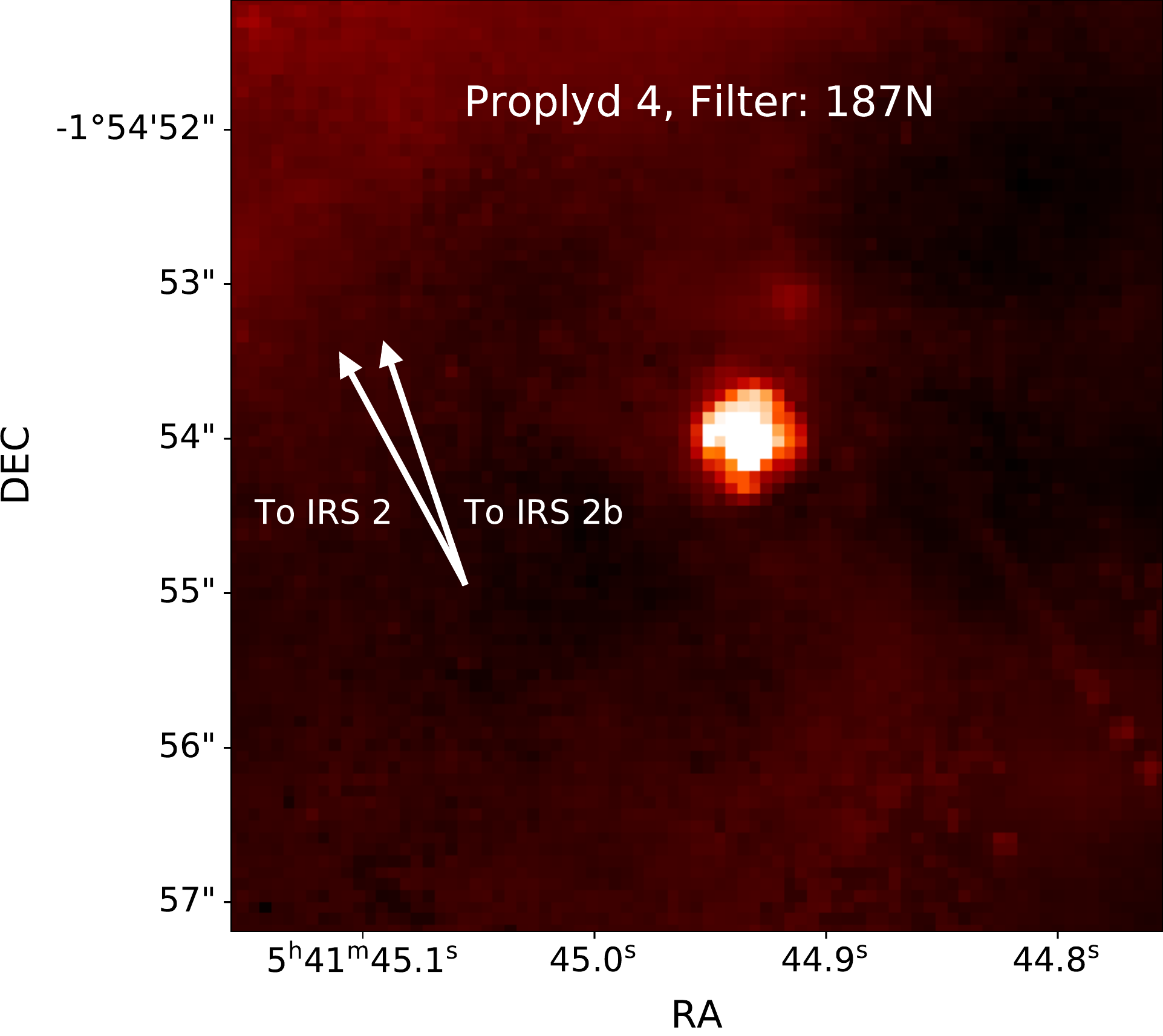}
    \includegraphics[width=5.86cm]{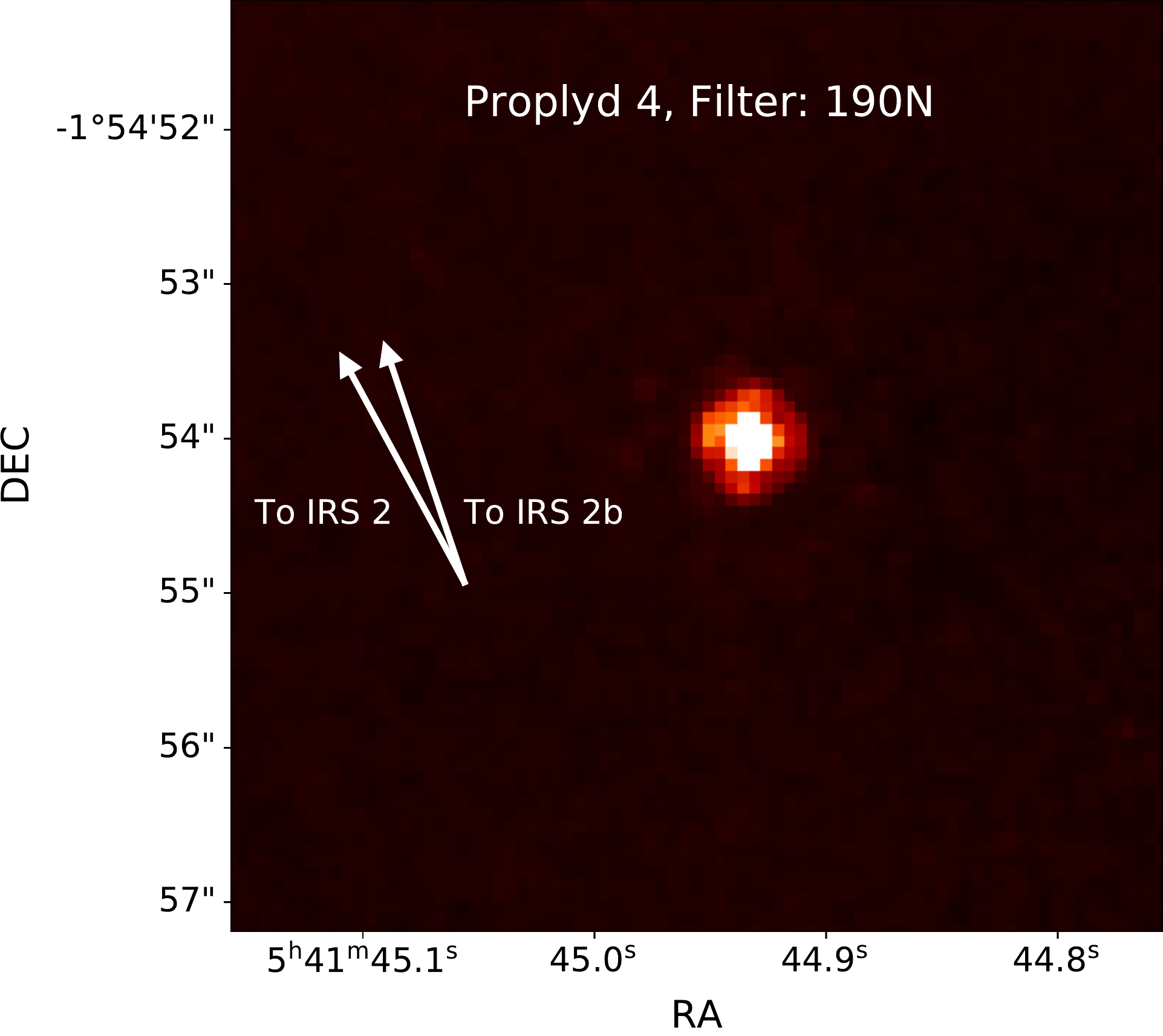}
    \includegraphics[width=5.86cm]{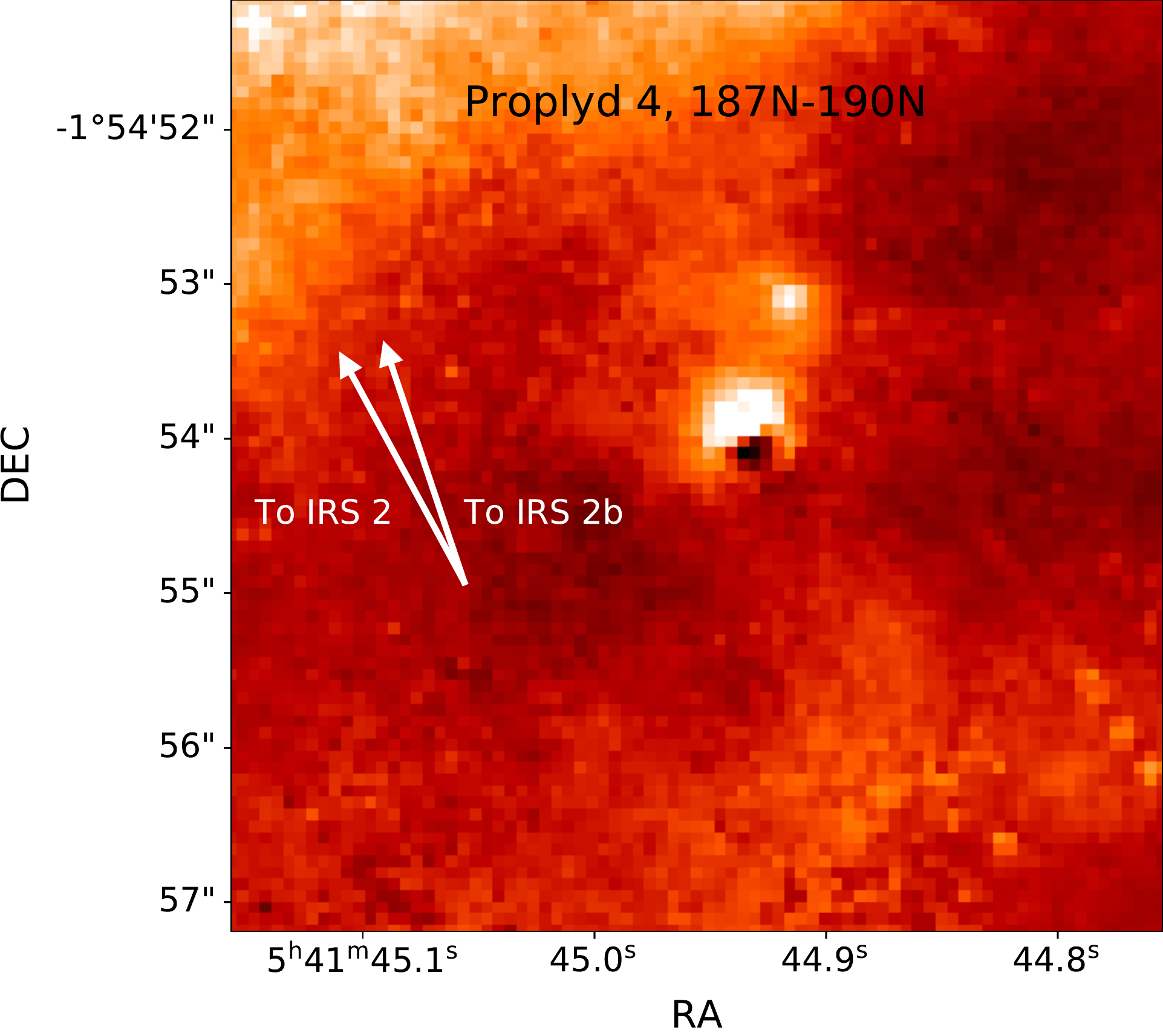}       
    \caption{A summary of the main proplyds introduced in this paper that were detected through Paschen\,$\alpha$ emission. The left hand panels are the Hubble F187N filter (centred on the Paschen\,$\alpha$ line) central panels are the F190N filter (the continuum) and right is F187N$-$F190 which is a proxy for the continuum subtracted Paschen\,$\alpha$ line emission. In each case the arrow points towards what is believed to be the UV source responsible for the photoevaporation. Note that Proplyd 2 may have a bipolar outflow, which is responsible for the elongated emission on either side. {Note that the epoch is J2000.}}
    \label{fig:Proplyds2to4}
\end{figure*}

Proplyd 4 was discovered in F187N-F190N (Paschen\,$\alpha$) emission. The F187N, F190N (continuum) and F187N-F190N images for Proplyd 4 are shown in the bottom row of panels in Figure \ref{fig:Proplyds2to4}. Proplyd 4 points towards IRS 2b, which lies at a projected distance of 0.054\,pc. An arrow is included on each panel in the direction of IRS 2b. 

Proplyd 4 exhibits a clear bright cusp with only a faint cometary tail. There is also an intriguing secondary Paschen\,$\alpha$ source which may well be an evaporating binary companion. The projected separation between these two objects is only about an arcsecond, corresponding to 413\,au at a distance of 414\,pc. However given that the orbital configuration is unknown 413\,au represents a lower limit on the true binary separation. Other binary proplyds have been discovered and, depending on the binary separation there may or may not be an interproplyd shell formed from the collision of the individual proplyd outflows \citep[e.g.][]{2002ApJ...570..222G, 2002RMxAA..38...71H, 2009ApJ...701L.100H, 2010RMxAA..46...79V, 2013ApJ...774...45W, 2018ApJ...854..144W}. In numerical simulations of evaporating binary propylds, \cite{2010RMxAA..46...79V} found that for separations $\sim2000$\,au an interproplyd shell does develop, but this is not the case for binaries of separation less than around 200\,au. With a separation of at least 400\,au in the case of Proplyd 4, it is conceivable that there might be an interproplyd shell, however it is not easily discerned in the archival Paschen\,$\alpha$ data.% {A new generation of observations of well known proplyds is in preparation with VLT/MUSE (Manara et al., in preparation) which may help with the }

%Follow up observations, e.g. with {VLT/MUSE}, will be required to confirm this one way or the other. 

From equations \ref{equn:Nly1} and \ref{equn:FUV1} the UV and ionising flux is $6.1\times10^6$\,$G_0$ and $1.6\times10^{13}$\,s$^{-1}$ respectively. From an approximate manual measurement of the radial extent of the cusp we assume an ionisation front radius of $87\,\pm37$\,au. The expected mass loss rate from equation \ref{eq:RIF} for this proplyd is hence $2.7^{+1.9}_{-1.5}\times10^{-7}$\,M$_\odot$\,yr$^{-1}$. 

Proplyd 4 also has a counterpart in \cite{vanTerwisga20}, where a dust mass of $26.8 \pm 0.87$\,M$_\oplus$ is inferred.

\subsection{Candidate Proplyd 5}
We now turn our attention to the more subtle, candidate, proplyds discovered in NGC2024. Although less clear, these all exhibit a proplyd morphology in the direction of either IRS 1 or IRS 2b. The candidate proplyds are also at larger distances, which probably explains why the proplyd morphology is less obvious. 

Candidate Proplyd 5 was discovered in F187N-F190N (Paschen\,$\alpha$) emission. A F187N-F190N image of Proplyd 5 is shown in the upper left panel of Figure \ref{fig:Proplyds5To8}. Candidate Proplyd 5 points towards IRS 2b, which lies at a projected distance of 0.124\,pc. An arrow is included in the direction of IRS 2b. From equations \ref{equn:Nly1} and \ref{equn:FUV1} the UV and ionising flux is $1.2\times10^6$\,$G_0$ and $3.1\times10^{12}$\,s$^{-1}$ respectively.

Candidate Proplyd 5 is very large radius cusp of $171^{+78}_{-31}$\,au, which corresponds to a mass loss rate of $3.2^{+5.7}_{-2.4}\times10^{-7}$\,M$_\odot$\,yr$^{-1}$ from equation \ref{eq:RIF}.  

Candidate Proplyd 5 also has a counterpart in \cite{vanTerwisga20}, where a dust mass of $40.4\pm0.71$\,M$_\oplus$ is inferred.

\subsection{Candidate Proplyd 6}
Candidate Proplyd 6 was discovered in F187N-F190N (Paschen\,$\alpha$) emission. A F187N-F190N image of Proplyd 6 is shown in the upper right panel of Figure \ref{fig:Proplyds5To8}. Candidate Proplyd 6 points towards IRS 2b, which lies at a projected distance of 0.11\,pc. An arrow is included in the direction of IRS 2b. From equations \ref{equn:Nly1} and \ref{equn:FUV1} the UV and ionising flux is $1.5\times10^6$\,$G_0$ and $3.9\times10^{12}$\,s$^{-1}$ respectively.

Candidate Proplyd 6 has only a faint cusp and was primarily identified through its cometary tail. 

From an approximate manual measurement of the radial extent of the cusp we assume an ionisation front radius of $77\pm31$\,au. This corresponds to a mass loss rate of $1.1^{+0.7}_{-0.6}\times10^{-7}$\,M$_\odot$\,yr$^{-1}$ from equation \ref{eq:RIF}.  

We could find no obvious counterpart for candidate proplyd 6 \cite{vanTerwisga20}

\subsection{Candidate Proplyd 7}
\begin{figure*}
    \centering
    \includegraphics[width=8.1cm]{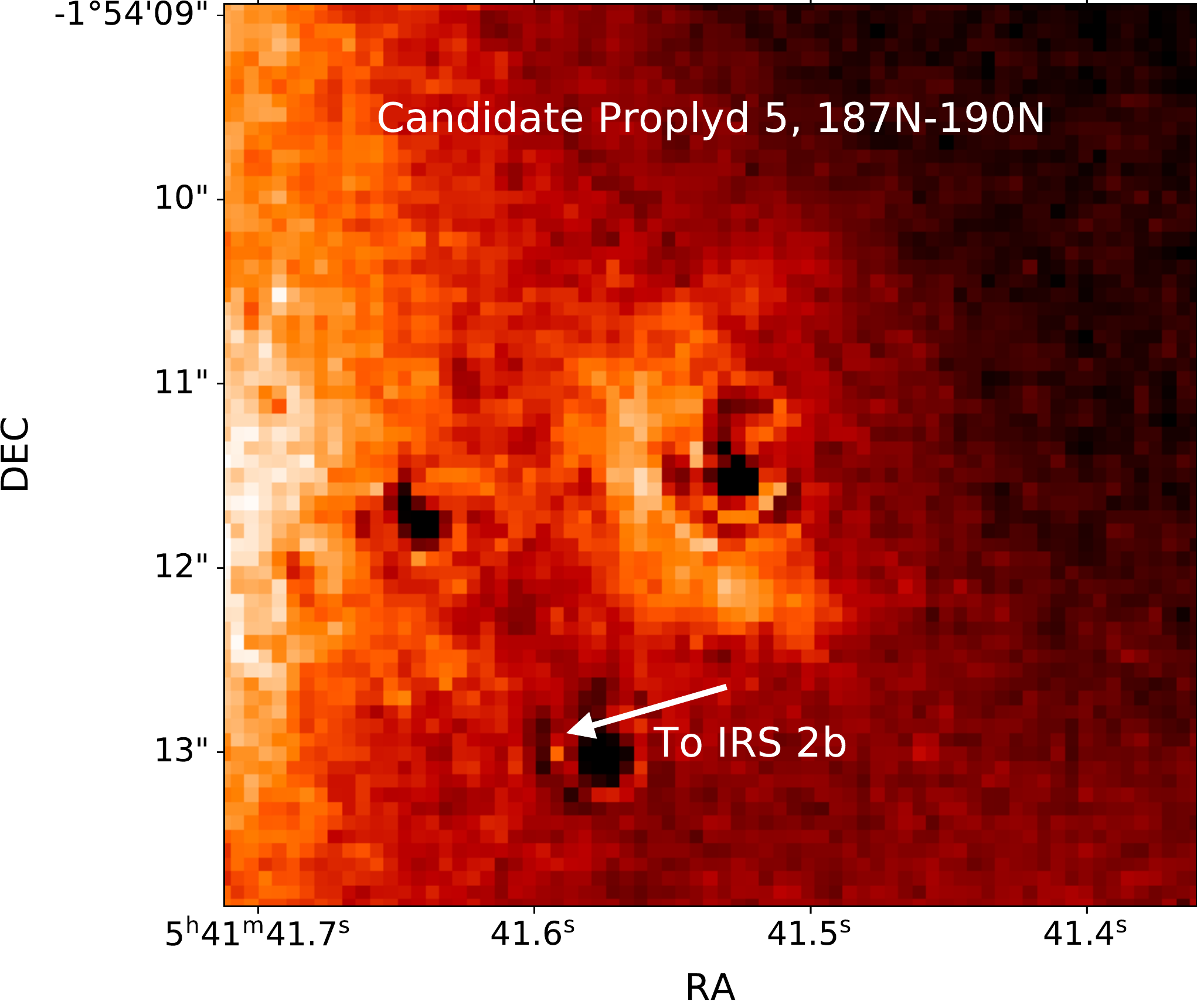}
    \includegraphics[width=8.25cm]{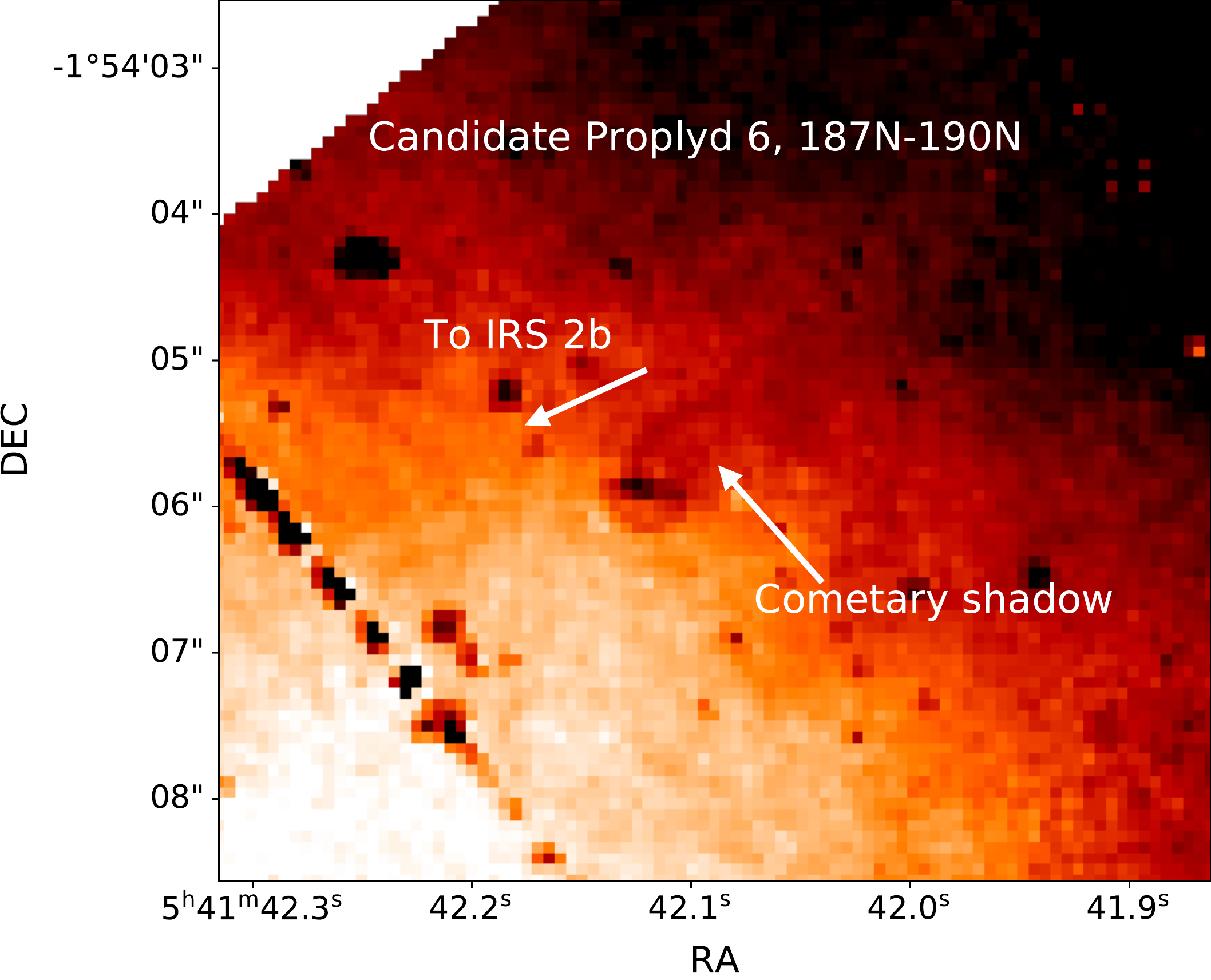}
    
    \hspace{-0.1cm}
    \includegraphics[width=8.2cm]{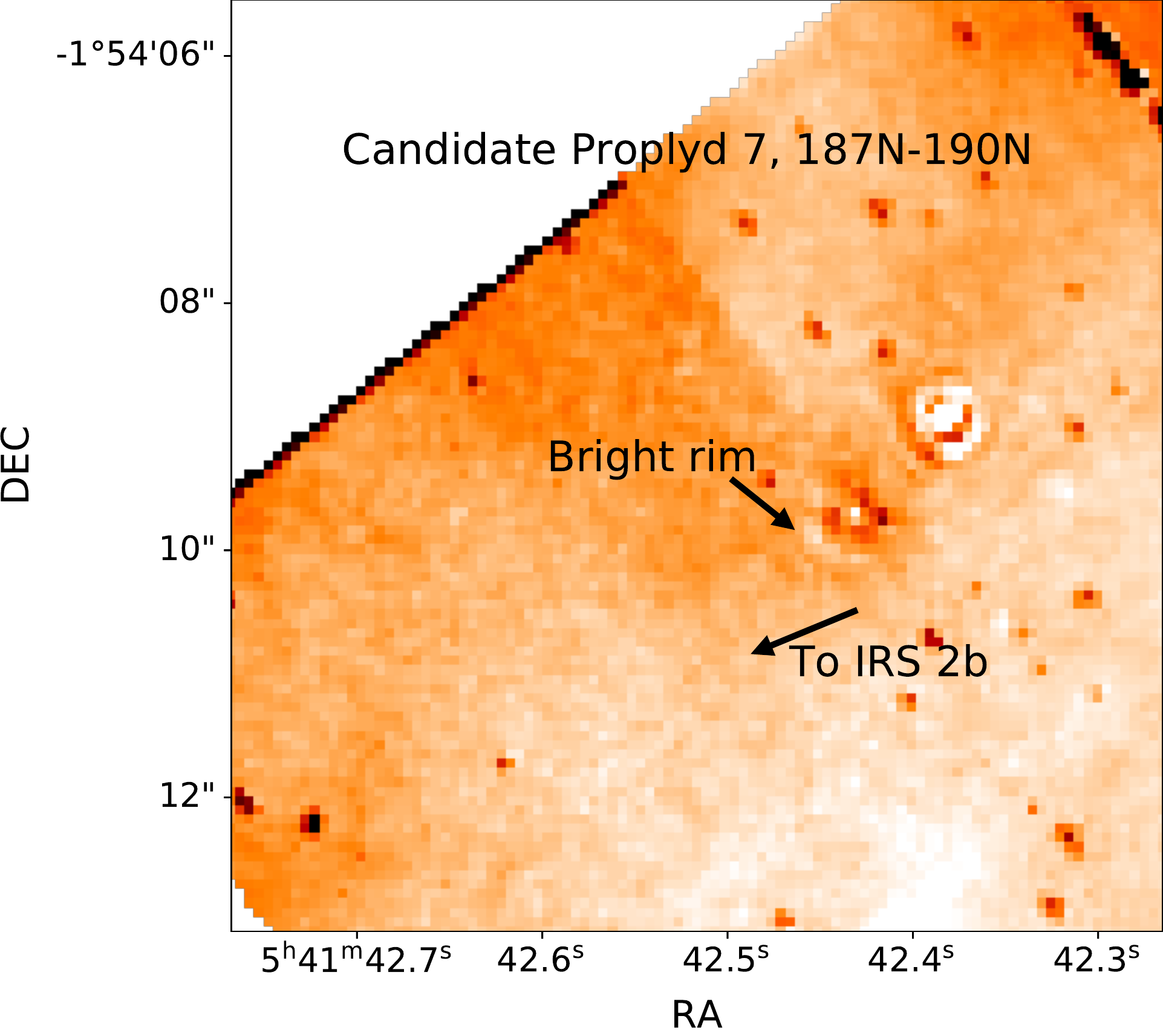}
    \includegraphics[width=8.3cm]{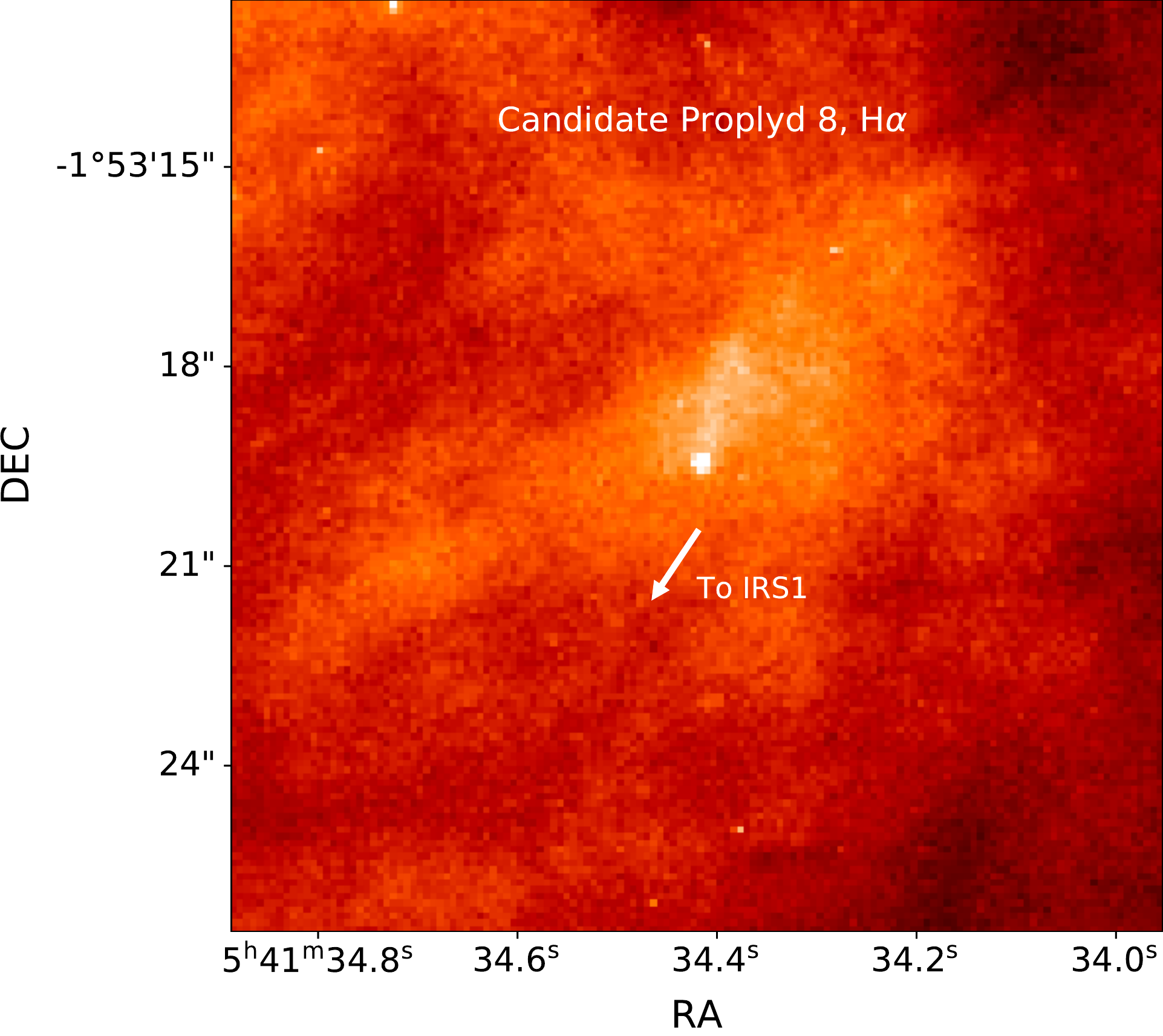}
    \caption{Additional candidate proplyds in NGC 2024, which are fainter or less obvious than the main proplyds introduced here. Candidates 5-7, detected in Paschen\,$\alpha$ are all close projected proximity to one another and pointing towards IRS 2b. Candidate 8 is detected in H\,$\alpha$ and pointing towards IRS 1. These candidate proplyds are typically at least a factor 2 more distant from the UV sources than the unsubtle proplyds in the region. {Note that the epoch is J2000.}}
    \label{fig:Proplyds5To8}
\end{figure*}

Candidate Proplyd 7 was discovered in F187N-F190N (Paschen\,$\alpha$) emission. A F187N-F190N image of Proplyd 7 is shown in the lower left panel of Figure \ref{fig:Proplyds5To8}. Candidate Proplyd 7 points towards IRS 2b, which lies at a projected distance of 0.10\,pc. An arrow is included in the direction of IRS 2b. From equations \ref{equn:Nly1} and \ref{equn:FUV1} the UV and ionising flux is $1.8\times10^6$\,$G_0$ and $4.7\times10^{12}$\,s$^{-1}$ respectively.

Candidate Proplyd 7 is close to candidates 5 and 6 and exhibits a cusp, but no obvious cometary tail. There is a bright Paschen\,$\alpha$ source slightly further than candidate 7 relative to IRS 2b. This is not a result of bad subtraction, but rather that this is a very bright Paschen\,$\alpha$ emitter, probably due to strong accretion. 

From an approximate manual measurement of the radial extent of the cusp we assume an ionisation front radius of $109\pm31$\,au. This corresponds to a mass loss rate of $2.0^{+0.9}_{-0.8}\times10^{-7}$\,M$_\odot$\,yr$^{-1}$ from equation \ref{eq:RIF}.

Candidate Proplyd 7 also has a counterpart in \cite{vanTerwisga20}, where a dust mass of $42.5\pm1.79$\,M$_\oplus$ was inferred.

\subsection{Candidate Proplyd 8}

Candidate Proplyd 8 was discovered in F656N (H\,$\alpha$) emission. A H\,$\alpha$ image of Proplyd 8 is shown in the lower right panel of Figure \ref{fig:Proplyds5To8}. Candidate Proplyd 8 points towards IRS 1, which lies at a projected distance of 0.19\,pc. An arrow is included in the direction of IRS 1. It is also possible that the tail of candidate proplyd 8 is a (very asymmetric) jet, though the alignment of the tail with the vector from IRS 1 is supportive of it being external photoevaporation. However, it is interesting to note that at least 2/3 of the proplyds in the older western region near IRS 1 have possible jets whereas none of the proplyds towards IRS 2b appear to. 

From equations \ref{equn:Nly1} and \ref{equn:FUV1} the UV and ionising flux is $2.0\times10^5$\,$G_0$ and $1.2\times10^{11}$\,s$^{-1}$ respectively. Candidate Proplyd 8 is in the weakest UV environment of the proplyds introduced in this paper. 

Candidate Proplyd 8 has a very small unresolved cusp (it spans a maximum of about 3 pixels in diameter) and a bright cometary tail. We assume an upper limit on the ionisation front radius of 60\,au. This corresponds to a mass loss rate of $<1\times10^{-8}$\,M$_\odot$\,yr$^{-1}$ from equation \ref{eq:RIF}.  

Candidate Proplyd 8 has no obvious counterpart in \cite{vanTerwisga20} and may be beyond the field they studied (it is beyond the limits of their Figure 1).

\section{Discussion}
\label{sec:discussion}

\begin{figure}
    \centering
    \vspace{-0.1cm}
    \includegraphics[width=9.5cm]{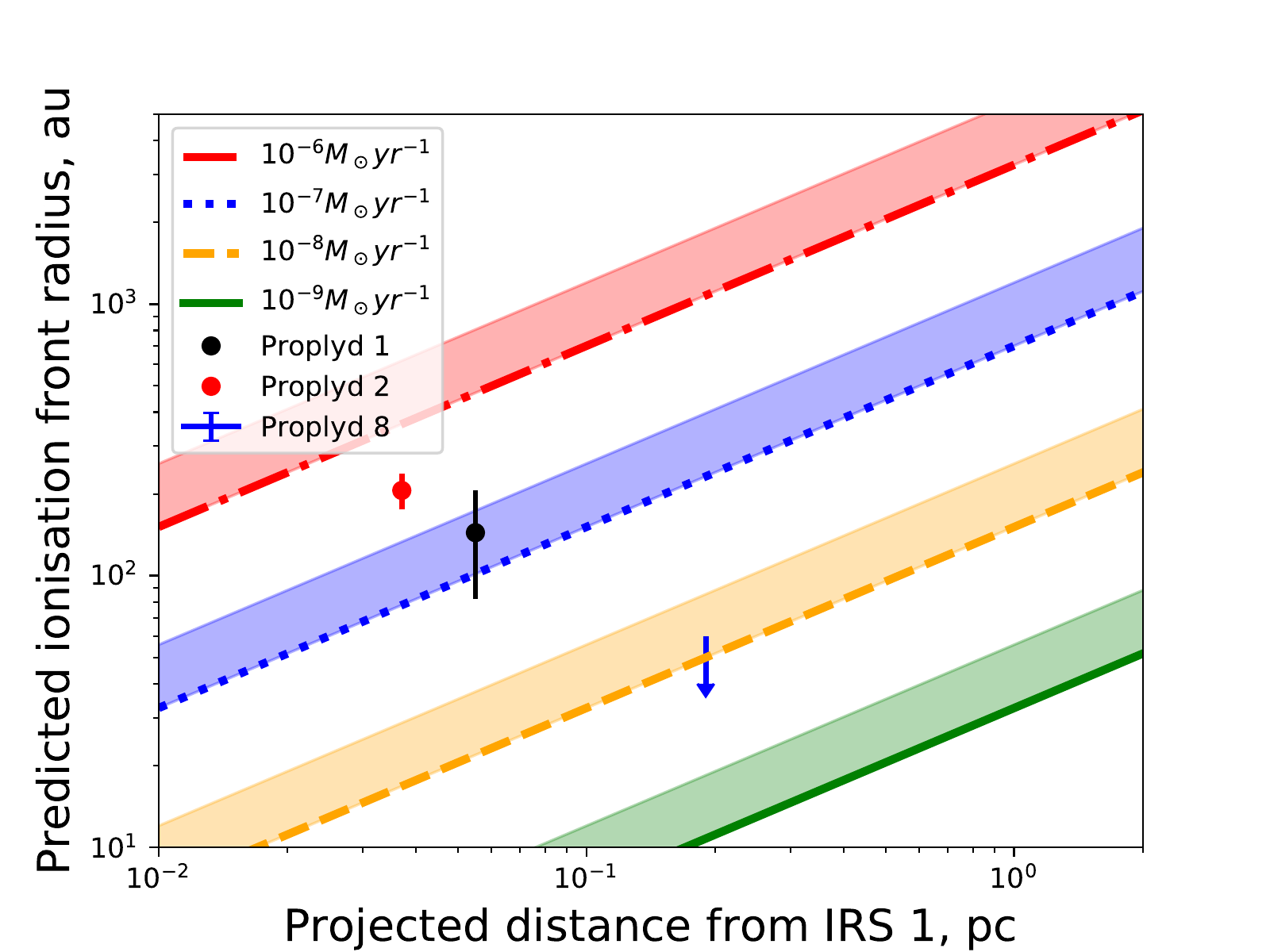}
    \includegraphics[width=9.5cm]{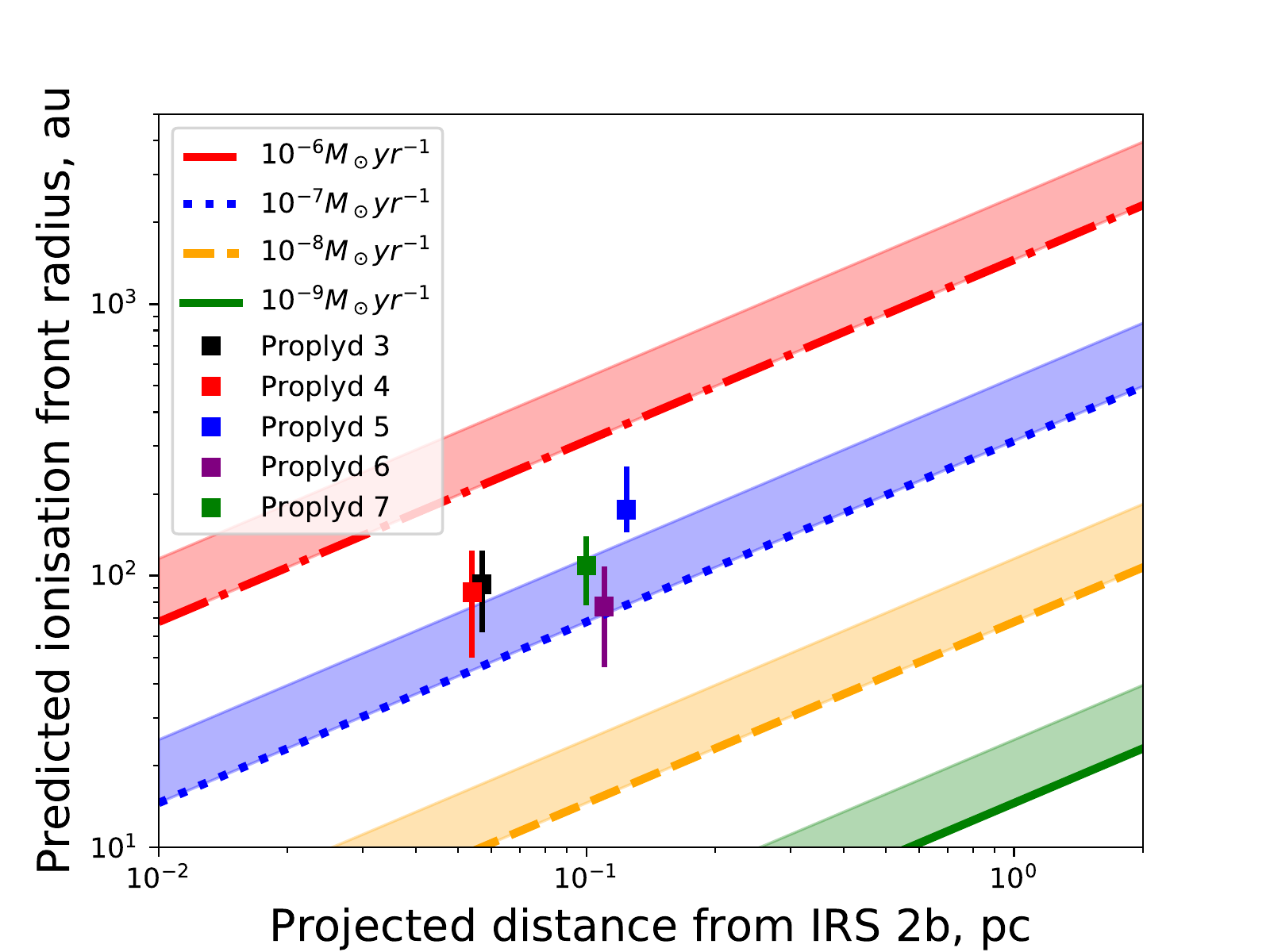}
    \caption{The predicted ionisation front radius for a proplyd as a function of the separation fron IRS 1 (upper panel) and IRS 2b (lower panel). The lines are our nominal values based on the ionising fluxes of \protect\cite{2003ApJ...599.1333S} and the shaded area goes out to ionising fluxes a factor 5 lower. In the upper panel the points are the 3 new proplyds in the vicinity of IRS 1 and the squares in the lower panel are new proplyds discovered in the vicinity of IRS 2b.   }
    \label{fig:RIF}
\end{figure}

\subsection{Mass loss rates and proplyd lifetimes}
\label{sec:Mdots}
Figure \ref{fig:RIF} shows the equation \ref{eq:RIF} predicted ionisation-front radius of a proplyd as a function of distance from IRS 1 (upper panel) and IRS 2b (lower panel). Different lines correspond to mass loss rates from $10^{-9}-10^{-6}$\,M$_\odot$\,yr$^{-1}$ using the UV fluxes from \cite{2003ApJ...599.1333S}. The shaded areas span a UV flux down to a factor 5 lower.  Overlaid are the ionisation front radii and projected separations of our observed proplyds on the panel of the suspected UV source responsible for photoevaporation (note that the estimated mass loss rates are all stated in section \ref{sec:proplyds} / Table \ref{tab:ProplydParams}).

The NGC 2024 proplyds (known so far) are all within 0.2\,pc of the exciting UV sources. The mass loss rates of proplyds near IRS 2b are generally higher than those near IRS 1. This is at least in part due to the stronger UV flux from the higher mass IRS 2b. However the discs near IRS 2b are also younger and therefore likely to be more massive and possibly also more extended which would also drive up the mass loss rate \citep[e.g.][]{2018MNRAS.475.5460H,2018MNRAS.481..452H,2020MNRAS.497L..40W}. 

The mass loss rates are generally higher than those inferred for the NGC 1977 proplyds ($\sim10^{-8}$\,M$_\odot$\,yr$^{-1}$) by \cite{Kim16}, with the exception of candidate proplyd 8. This is because the NGC 2024 proplyds are closer to the UV sources (meaning stronger UV fluxes) as well as younger and hence possibly more massive/extended. 

The ratio of a disc mass estimate to the mass loss rate gives a zeroth order estimate of the disc lifetime, however this comes with a number of caveats. The mass loss rate is typically a strong function of disc radius \citep[e.g.][]{2018MNRAS.481..452H} and an evaporating disc gets truncated over time. High mass loss rates hence quickly result in a smaller disc with a much lower mass loss rate that survives much longer than one would anticipate based on the earlier mass loss rate. In addition, the UV flux that any given disc is exposed to at $<0.1$\,pc distances from the UV source for $\sim$km\,s$^{-1}$ velocity dispersions can change by up to two orders of magnitude within 1\,Myr \citep{1999ApJ...515..669S, 2019MNRAS.490.5478W}. {We also estimated the mass loss rate using the projected separation, when the true separation could be somewhat larger (and hence mass loss rate lower)}. The other key caveat here is that the \cite{vanTerwisga20} masses are in the {mm} dust only and it is the gas that is predominantly evaporated \citep{2016MNRAS.457.3593F, 2020MNRAS.492.1279S}. These caveats aside, assuming a dust-to-gas mass ratio of 1/100 the proplyd $M/\dot{M}$ values are typically only a few tens of kyr, with the largest value being $\sim120$\,kyr for Proplyd 1. For the younger proplyds near IRS 2b the argument could be made that they have only recently begun being subject to photoevaporation (e.g. as IRS 2b ceased being so embedded along certain lines of sight). However proplyd 1, near IRS 1, is in the $\sim1$\,Myr old region, raising the proplyd lifetime problem known from the ONC \citep[e.g.][]{1994ApJ...436..194O}. The solution to this seems to require either a lower dust-to-gas mass ratio \citep[and hence higher gas mass than anticipated from continuum measurements,][]{2020MNRAS.492.1279S}, ongoing star formation throughout NGC 2024 (even in the depleted western region) or that the remaining proplyds in the west recently migrated in towards IRS 1 \citep{2019MNRAS.490.5478W}. The former solution of higher-than-canonical mass discs was also required by \cite{2018MNRAS.475.5460H} to explain the formation of Trappist-1, even in weak UV environments and high mass discs have since been shown to be sustainable in a gravitationally stable manner around low mass stars \citep{2020MNRAS.494.4130H, 2020MNRAS.492.5041C}. 

It is also important to note that proplyds have now been found in regions with ages ranging from $<0.5$\,Myr (this paper) to around 1.5\,Myr \citep{Kim16, 2018AJ....156...84K}, so higher gas masses, ongoing star formation, or continuous migration of discs into the high UV environment would have to be taking place to observe proplyds over such a length of time.

\subsection{Indirect evidence for external photoevaporation?}
Much recent evidence for disc evaporation has come indirectly from the distribution of disc properties in a cluster, for example searching for correlations between the disc mass, radius or disc fraction as a function of projected separation from a UV source \citep[e.g.][]{2014ApJ...784...82M, 2016arXiv160501773G, 2017AJ....153..240A, 2018ApJ...860...77E}. \cite{2000AJ....120.1396H} found no obvious trend in the {infrared} disc fraction as a function of projected radius in NGC 2024, but we  looked for any imprint of radiation environment on the \cite{vanTerwisga20} disc \textit{masses}. 

NGC 2024 is a challenging region to look at in a statistical sense as proplyds are associated with both IRS 1 and IRS 2b. There is also a lot of gas and dust in the  eastern region. The UV structure of the region is hence likely set by a complicated spatially varying combination of the two stars. 

\begin{figure}
    \centering
    \includegraphics[width=9.5cm]{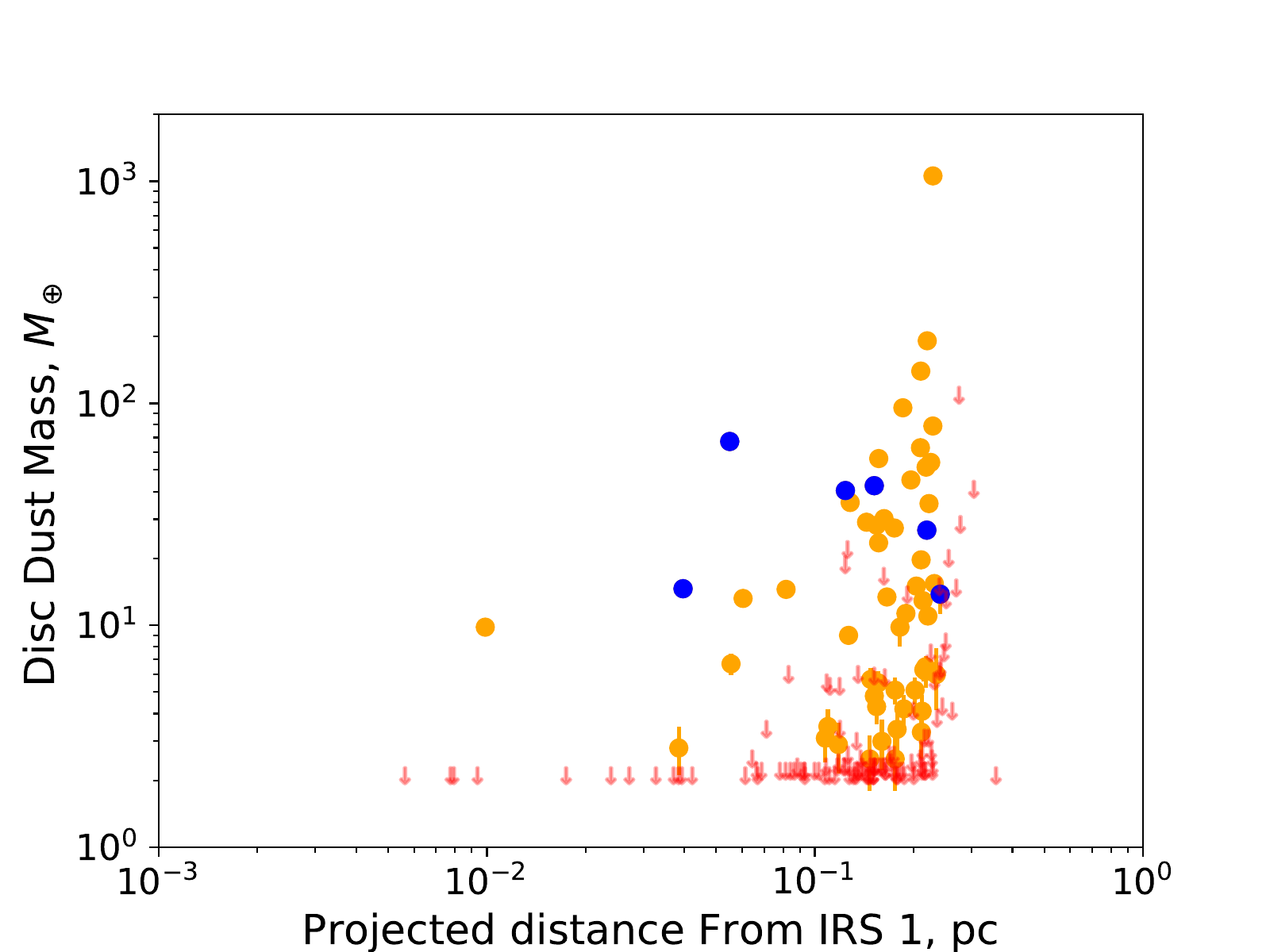}
    \includegraphics[width=9.5cm]{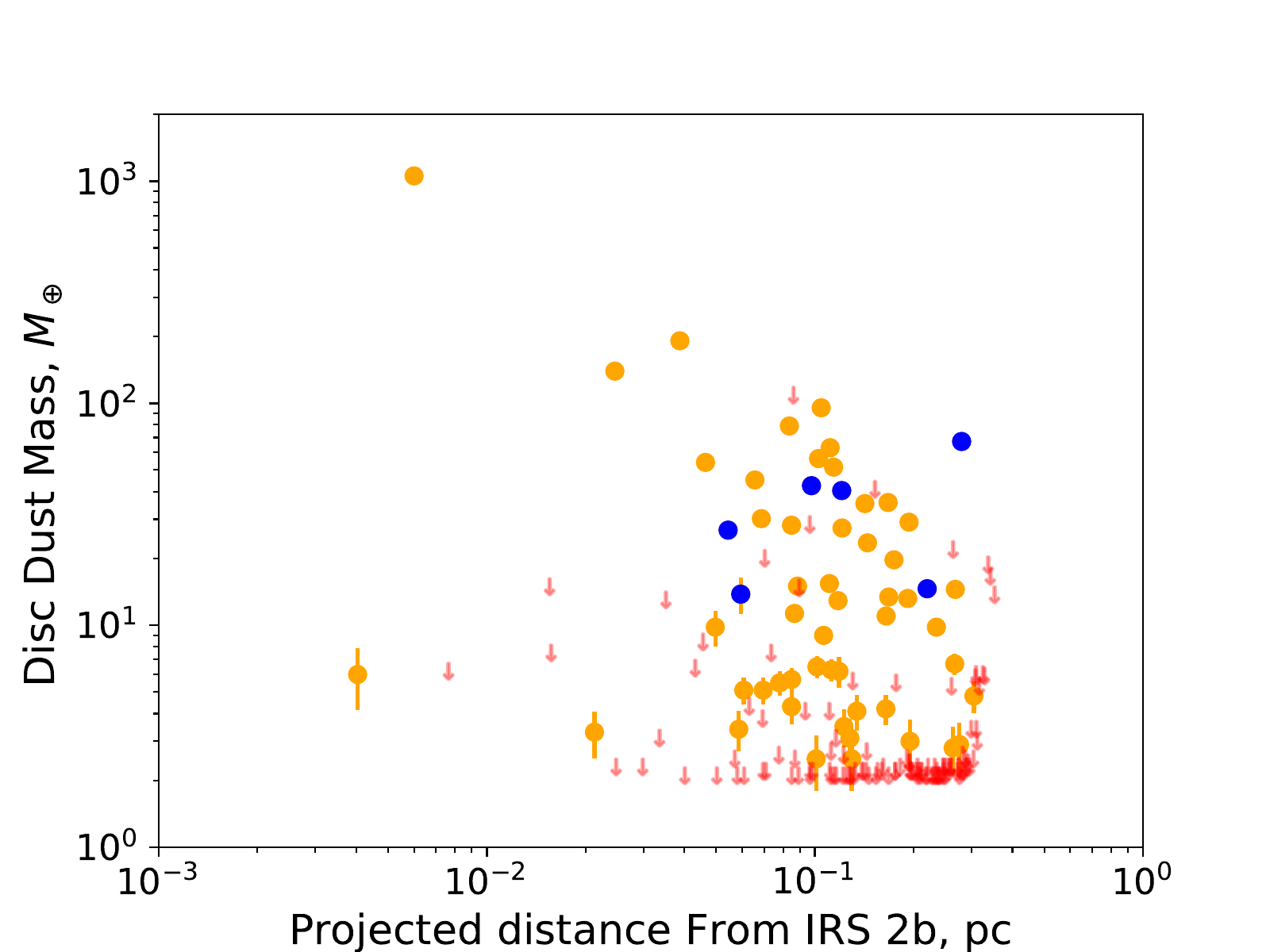}    
    \caption{Disc mass as a function of projected separation from IRS 1 (upper panel) and IRS 2b (lower panel) for the discs with mass estimates from \protect\cite{vanTerwisga20}. The blue points are the proplyds introduced in this paper for which we could find \protect\cite{vanTerwisga20} counterparts. The red arrows are the discs for which \protect\cite{vanTerwisga20} have upper limits.} 
    \label{fig:M_dist}
\end{figure}

Figure \ref{fig:M_dist} shows the \cite{vanTerwisga20} disc dust mass estimates as a function of projected separation from IRS 1 and IRS 2b. The blue points represent the new proplyds introduced in this paper (where mass estimates are available) and the red arrows are the discs for which \cite{vanTerwisga20} could only place upper limits on the dust mass. We evaluate the Pearson's measure of linear correlation and Spearman's measure of higher order correlation. In each case we also evaluate the p-value by randomly shuffling the arrays of projected separation/disc mass ten thousand times, re-computing the correlation metrics and keeping track of the fraction of distributions with weaker metrics than the true one. 

In Figure \ref{fig:M_dist}  we include all discs with  \cite{vanTerwisga20} mass constraints in each panel, regardless of whether or not they are closer in projected separation to IRS 1 or IRS 2b. Disc masses are found to weakly increase with increasing distance from IRS 1 (Pearson's 0.2, p-value 0.12) (Spearman's 0.27, p-value 0.05) though this will be influenced by the fact that there is a distinct population of younger more massive discs in the vicinity of IRS 2b. The disc masses are also found to weakly decrease with projected separation from IRS 2b (Pearson's -0.27, p-value 0.05), (Spearman's -0.2, p-value 0.13). The proplyds do not obviously stand out from the rest of the population, other than that they are some of the more massive members of the group (which gives a higher mass loss rate). %This lends further weight to the suggestion above that the proplyds here either migrated in or formed  relatively recently to circumvent the proplyd lifetime problem \citep{2019MNRAS.490.5478W}. 

\begin{figure}
    \centering
    \includegraphics[width=9.5cm]{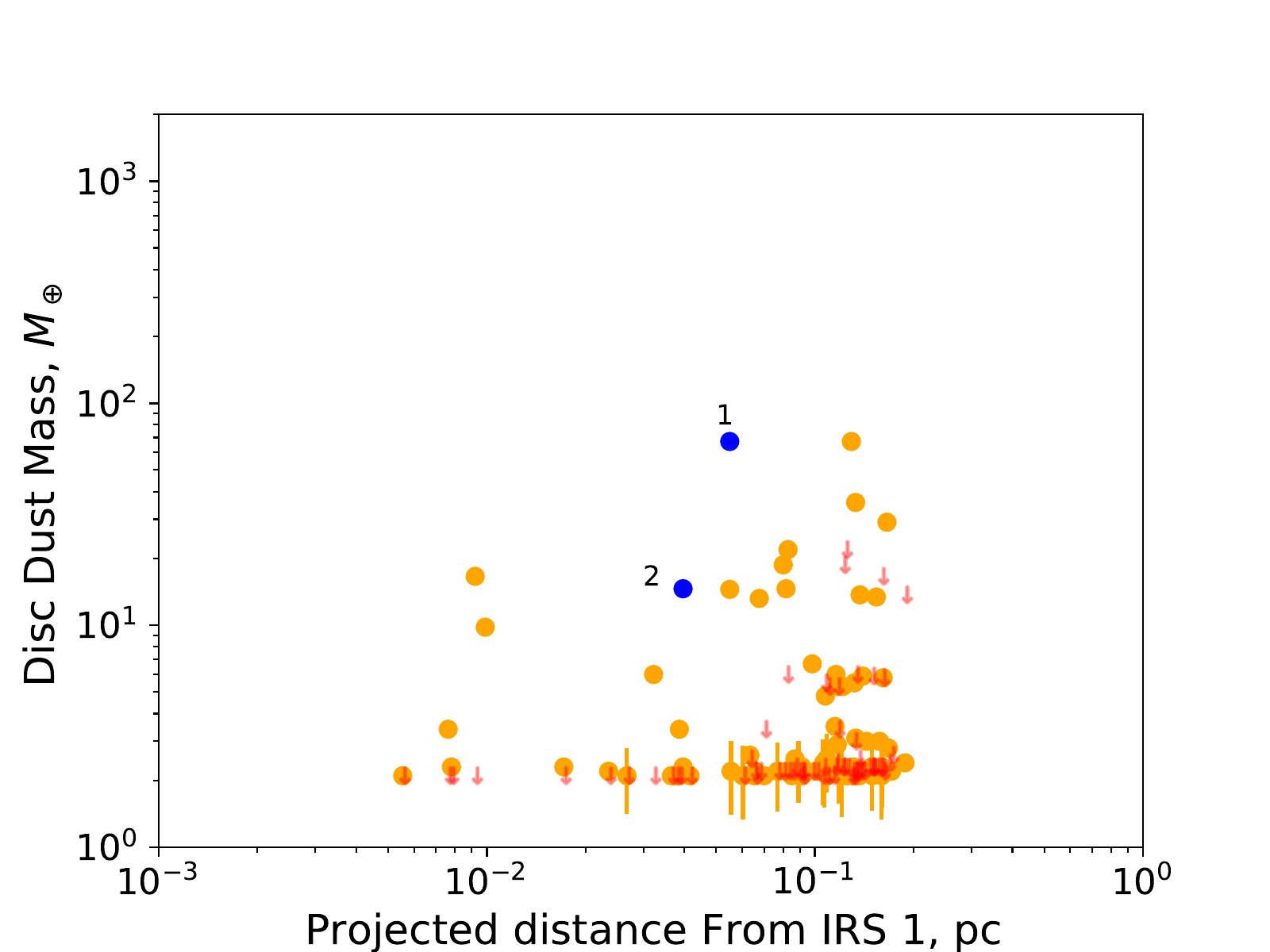}
    \includegraphics[width=9.5cm]{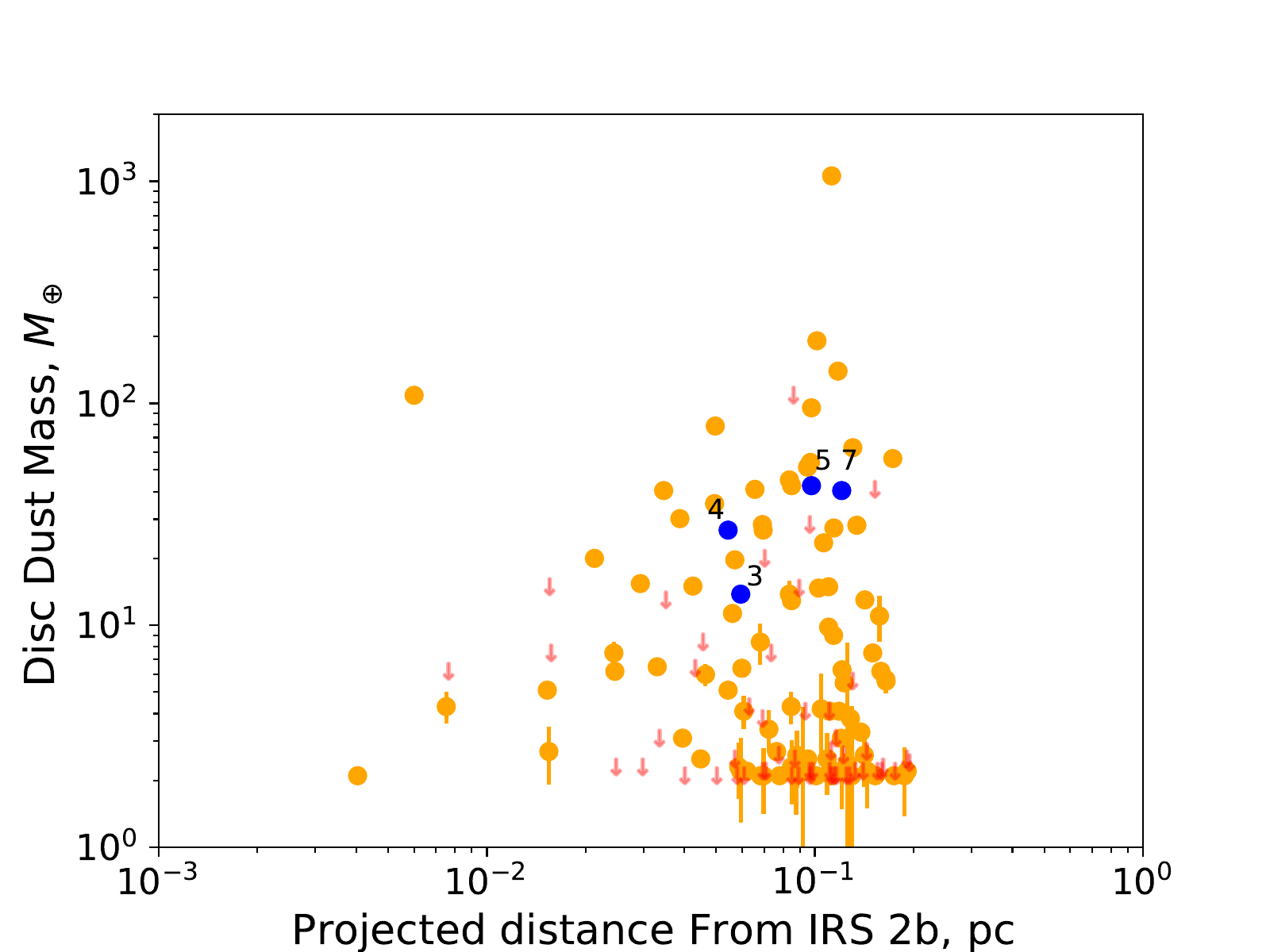}    
    \caption{As in Figure \protect\ref{fig:M_dist}, these plots show the disc dust mass as a function of projected separation from IRS 1 or IRS 2b. The difference here is that we only include the discs nearest to IRS 1 (upper panel) or IRS 2b (lower panel) iin projected separation, whereas in Figure \protect\ref{fig:M_dist} all discs are included in both panels. The {numbered} blue points are the proplyds introduced in this paper for which we could find \protect\cite{vanTerwisga20} counterparts. The red arrows are the discs for which \protect\cite{vanTerwisga20} have upper limits. }
    \label{fig:M_dist_filter}
\end{figure}

In Figure \ref{fig:M_dist_filter} we again plot the disc mass as a function of projected separation, but filtering the data such that only the discs nearest to the UV source in projected separation are included. Correlation factors for IRS 1 are now (Pearson's 0.028, p-value 0.81) (Spearman's 0.39, p-value $2\times10^{-4}$) and for IRS 2b (Pearson's -0.23, p-value 0.013) (Spearman's -0.144, p-value 0.16).  Most of the upper limits correspond to the same disc mass, but re-computing after discarding all upper limits less than 2.5\,M$_\oplus$ only makes the Pearson's and Spearman's correlations even weaker. 
So there is still no strong correlation with disc properties as a function of projected distance from either source. 

A key point to explain the lack of \textit{spatially varying} environmental imprint on the disc statistics is that the field considered here is rather small, with all discs being within $\sim 0.3$\,pc of the UV sources. The correlation between disc mass and projected separation from $\sigma$ Ori discovered by \cite{2017AJ....153..240A} only kicks in beyond around 0.5\,pc, with a flat distribution of masses interior to that.  Over such a small distance scale the crossing timescale for a $\sim1$\,km\,s$^{-1}$ velocity dispersion is $\sim0.1\,$Myr. Since both the East and Western zones are many crossing timescales old the disc masses would be randomised with projected separation.

In the older western region a larger fraction of discs only have upper limits on the mass, consistent with the picture that this is older and more of the discs have been heavily depleted by photoevaporation. In the younger eastern region around IRS 2b, many discs are younger and are probably still well shielded by the immediate interstellar medium. It is also worth noting that all of the proplyds associated with IRS 2b are at the edge of the dense ridge, which would be consistent with the young stars/discs being progressively exposed to the UV as the gas that they are embedded in is dispersed. So even if the region were not old enough for the crossing time to wash out any spatially varying imprint of evaporation, there may be no signature of evaporation as a function of projected separation simply because the discs are young and embedded.

\subsection{External photoevaporation in NGC 2024: reflecting upon the van Terwisga et al. (2020) interpretation}
\cite{vanTerwisga20} presented ALMA 1.3mm continuum observations of the two distinct populations in NGC 2024. They suggested that a key distinction was that the  {mm}-depleted western population is much less embedded and possibly subject to evaporation by IRS 1 and IRS 2b. They also suggested that the discs in the western population may have just formed smaller and hence had shorter viscous lifetimes. It is important to understand if it is possible for there to be systematic regional variations in disc properties independent of environmental influence like photoevaporation (e.g. set by the initial molecular cloud properties there). One possible mechanism to enable this would be variations in the cosmic ray ionisation rate, which \cite{2020arXiv200600019K} showed can affect the magnetorotational braking during collapse and lead to smaller disc sizes. 

Our discovery of proplyds in NGC 2024 has some important implications for the discussion above. First of all, we have indeed confirmed that external photoevaporation is at work in the region. In the western region our proplyds point towards IRS 1, implying that the IRS 1 dominates the UV radiation field there and could have depleted the disc fraction in the region even without the help of IRS 2b.  

What was not anticipated from \cite{vanTerwisga20} is ongoing photoevaporation in the eastern population, but with 5 of our proplyds in this zone we have shown that this is actually the case. {Despite the eastern region being younger and having a {higher} disc fraction in terms of ALMA detections ($45\pm7$\,per cent) compared to the western region, the disc fraction is still low compared mm continuum detection rate in other star forming regions, such as Lupus \citep[70\,per cent mm continuum detection fraction in a 1-2\,Myr region][]{2016ApJ...828...46A} and Sigma Ori \citep[40\,per cent mm continuum detection fraction in a 3-5\,Myr region][]{2017AJ....153..240A}. Though future comparisons ensuring equivalent detection thresholds will be necessary, this is evidence that many discs in the eastern region {of NGC 2024} that have not been shielded by the ISM have been subject to photoevaporation by IRS 2b.

\subsection{Implications for planet formation}
Our understanding of planet formation in the idealised scenario of discs as isolated systems is still not yet fully understood. By extension, the added complication of the impact of external photoevaporation on planet formation is not well understood. What is known is that external photoevaporation reduces the disc mass \citep[e.g.][]{2017AJ....153..240A, Winter18b, 2020MNRAS.492.1279S}, radius \citep[e.g][]{2017MNRAS.468L.108H, 2018ApJ...860...77E, 2020MNRAS.492.1279S} and lifetime \citep[e.g.][]{2016arXiv160501773G, 2019MNRAS.490.5678C, 2020MNRAS.491..903W} which could all of course indirectly affect the formation of planets. 

A key element for planet formation, even in the inner disc, is the dust reservoir of the \textit{entire} disc. In the outer disc grains grow and drift inwards to smaller radii and can contribute to planet formation there, which is expected to be necessary to form the large mass of close-in planets  \citep[e.g.][]{2012A&A...539A.148B, 2014ApJS..210...20M, Ormel17Trappist}. 

In the outer disc, small dust grains get entrained in the photoevaporative wind \citep{2016MNRAS.457.3593F}, whereas larger grains are pushed inwards with the retreating disc outer edge and could still ultimately contribute to planet formation \citep{2020MNRAS.492.1279S}. The total dust mass lost in a disc is set by the competition between grain growth in the disc (which is fastest at small orbital radii) and the rate of external photoevaporation \citep[e.g.][]{2018MNRAS.475.5460H, 2020MNRAS.492.1279S}. If grain growth proceeds through the disc quickly enough, then only a small amount of dust would end up entrained in the wind. Conversely very strong, early photoevaporation could deplete significant amounts of the potential dust reservoir that could otherwise have migrated in to contribute to inner planet formation. A key question is therefore how long discs remain embedded, permitting dust to grow to a size where it is immune to photoevaporation once the source becomes exposed to the ultraviolet source.

In light of the above, our finding evidence for very early photoevaporation throughout NGC 2024 has added importance because it may also be significantly reducing the dust mass reservoir (in addition to the usual expected impact on the gas disc mass/radius/lifetime). This would also explain why the dust mass estimates of discs in the $0.2-0.5$\,Myr eastern population of NGC 2024 are more comparable to those in $1-3$\,Myr old regions \citep{vanTerwisga20}.

{There is growing evidence for planet formation on $<1$\,Myr timescales \citep[e.g.][]{2020Natur.586..228S}, but here we show evidence for very early external photoevaporation (0.2-0.5\,Myr). }

\subsection{Potential for additional evaporating discs}
\label{sec:futureSearch}
Our search for externally photoevaporating discs in NGC 2024 using archival data is certainly not complete. The F187/F190 fields using NICMOS have a small field of view and the archival data does not cover the whole region. Future HST WFC3 $H\,\alpha$ and Paschen\,$\alpha$ surveys may reveal more proplyds in the central region. Additional searches at larger distances from the less embedded western region may also reveal photoevaporative winds from discs where the UV field is weaker, e.g. in CI \citep{2020MNRAS.492.5030H}. Surveys out to larger distances from IRS 1 and IRS 2b may also reveal any statistical imprint of photoevaporation on the disc properties. The above combined with the need to confirm our candidate proplyds means that NGC 2024 is a prime region to search for further evidence for external disc photoevaporation.

\section{Summary and conclusions}
\label{sec:conclusions}
NGC 2024 is a young star forming region in Orion with a radial age gradient \citep{2014ApJ...787..109G}. \cite{vanTerwisga20} also recently found two slightly differently aged {sub-}populations with very different ALMA 1.3\,mm continuum disc fractions in the inner ($<0.5$\,pc) part of the region. To gauge the role of external photoevaporation of discs in sculpting these populations we searched for proplyds in archival HST observations of NGC 2024. We draw the following main conclusions in this work. \\

\noindent 1) We discovered 4 firm proplyds and 4 candidate proplyds. Three of these are in the vicinity of the B0.5V star IRS 1 and the other 5 associated with the (younger and more embedded) O8V star IRS 2b. External photoevaporation of discs is definitely occurring in NGC 2024 \textit{and within both the statistically older and younger populations.} It is also worthy of note that together with NGC 1977 \citep{Kim16} early type B stars have now clearly been demonstrated as capable of evaporating the discs of nearby stars.  \\

\noindent 2) The geometry of the cometary proplyds introduced here can be explained using only IRS1 and IRS2b. No further embedded UV source is required. Proplyds 1 and 2 in the western region show evidence for a jet because of elongated emission through the cusp.  \\

\noindent 3) Proplyd 4 in this paper is possibly a binary proplyd with a separation of at least 400\,au, though we cannot discern an interproplyd shell with the archival Paschen\,$\alpha$ observations. Follow up observations of this system will be particularly interesting. \\

\noindent 4) With a maximum proplyd $M/\dot{M}$ of  $\sim120$\,kyr, even with the caveats of such an estimate there appears to be a proplyd lifetime problem in NGC 2024. Furthermore, the proplyds are actually some of the more massive discs from \cite{vanTerwisga20}. This suggests that either discs are more massive in gas than expected from continuum measurements \citep[i.e. that the dust-to-gas mass ratio is lower than $1/100$][]{2020MNRAS.492.1279S}, that there is ongoing migration of relatively massive discs into the high UV inner part of the cluster, or ongoing star formation, in both regions \citep{2019MNRAS.490.5478W}. \\

\noindent 5) There is no clear statistical imprint of evaporation on the disc masses in the region as a function of projected separation from the main UV sources, but this is unsurprising given the young age of the region and that the field being considered is very small so discs that are not embedded are evaporated to the same degree (i.e. rapidly). Furthermore, the dynamical time in this compact central region is short ($\sim0.1\,$Myr) so any imprint of photoevaporation with projected separation would also quickly be washed out. \\

\noindent 6) Our results slightly change the \cite{vanTerwisga20} interpretation of the region in a few ways. Firstly, evaporation of the depleted western population appears to be dominated by IRS 1, since the proplyds in the region are pointing towards it, and so IRS 2b is not required to play a role there. Second, the eastern population is also being actively evaporated by IRS 2b where discs are not embedded in the ISM, which may also explain why the {ALMA mm continuum} disc fraction {is more like that of an older region \citep{vanTerwisga20}}. \\

{\noindent 7) Our evidence for external photoevaporation in the 0.2-0.5\,Myr eastern region shows that this environmental influence can be in competition even with very early planet formation \citep[e.g.][]{2020Natur.586..228S}.} \\

\section*{Acknowledgements}
{We warmly thank the anonymous referee for their insightful review. }
TJH is funded by a Royal Society Dorothy Hodgkin Fellowship, which also supports GB.
JSK was partially supported by NASA under Agreement No. NNX15AD94G for the program  “Earths in 
Other Solar Systems.
AJW acknowledges funding from an Alexander von Humboldt Stiftung Postdoctoral Research Fellowship. ADS thanks the Science and Technology Facilities Council (STFC) for a Ph.D. studentship and CJC acknowledges support from the STFC consolidated grantST/S000623/1. This work has also been supported by the European Union’s Horizon 2020 research and innovation programme under the Marie Sklodowska-Curie grant agreement No 823823 (DUSTBUSTERS).

Support for this work for Stapelfeldt \& Hines was provided by 
NASA through Grant Number GO-9424 from STScI, operated by the 
Association of Universities for Research in Astronomy Incorporated, 
under NASA contract NAS5-26555.

This research made use of Astropy\footnote{\url{http://www.astropy.org}}, a community-developed core Python package for Astronomy \citep{astropy:2013,astropy:2018}

This work used observations made with the NASA/ESA Hubble Space Telescope. Some were obtained from the Hubble Legacy Archive, which is a collaboration between the Space Telescope Science Institute (STScI/NASA), the Space Telescope European Coordinating Facility (ST-ECF/ESA) and the Canadian Astronomy Data Centre (CADC/NRC/CSA). Some of the HST data presented in
this paper was obtained from the Mikulski Archive for Space
Telescopes (MAST) at \url{https://archive.stsci.edu/hst/}. STScI is
operated by the Association of Universities for Research in
Astronomy, Inc., under NASA contract NAS5-26555. Support
for MAST for non-HST data is provided by the NASA Office
of Space Science via grant NNX09AF08G and by other grants
and contracts.

Herschel is an ESA space observatory with science instruments provided by European-led Principal Investigator consortia and with important participation from NASA.

\section*{DATA AVAILABILITY}
The data here is all publicly available in the MAST science archive, the Hubble Legacy Archive (HLA) and the ESA Herschel Science Archive. We have made reference throughout the paper to the relevant PI's and observations.

\bibliographystyle{mnras}
\bibliography{references}{}

\begin{thebibliography}{}
\makeatletter
\relax
\def\mn@urlcharsother{\let\do\@makeother \do\$\do\&\do\#\do\^\do\_\do\%\do\~}
\def\mn@doi{\begingroup\mn@urlcharsother \@ifnextchar [ {\mn@doi@}
  {\mn@doi@[]}}
\def\mn@doi@[#1]#2{\def\@tempa{#1}\ifx\@tempa\@empty \href
  {http://dx.doi.org/#2} {doi:#2}\else \href {http://dx.doi.org/#2} {#1}\fi
  \endgroup}
\def\mn@eprint#1#2{\mn@eprint@#1:#2::\@nil}
\def\mn@eprint@arXiv#1{\href {http://arxiv.org/abs/#1} {{\tt arXiv:#1}}}
\def\mn@eprint@dblp#1{\href {http://dblp.uni-trier.de/rec/bibtex/#1.xml}
  {dblp:#1}}
\def\mn@eprint@#1:#2:#3:#4\@nil{\def\@tempa {#1}\def\@tempb {#2}\def\@tempc
  {#3}\ifx \@tempc \@empty \let \@tempc \@tempb \let \@tempb \@tempa \fi \ifx
  \@tempb \@empty \def\@tempb {arXiv}\fi \@ifundefined
  {mn@eprint@\@tempb}{\@tempb:\@tempc}{\expandafter \expandafter \csname
  mn@eprint@\@tempb\endcsname \expandafter{\@tempc}}}

\bibitem[\protect\citeauthoryear{{Adams}, {Hollenbach}, {Laughlin}  \&
  {Gorti}}{{Adams} et~al.}{2004}]{2004ApJ...611..360A}
{Adams} F.~C.,  {Hollenbach} D.,  {Laughlin} G.,   {Gorti} U.,  2004, \mn@doi
  [\apj] {10.1086/421989}, \href
  {http://adsabs.harvard.edu/abs/2004ApJ...611..360A} {611, 360}

\bibitem[\protect\citeauthoryear{{Ansdell} et~al.,}{{Ansdell}
  et~al.}{2016}]{2016ApJ...828...46A}
{Ansdell} M.,  et~al., 2016, \mn@doi [\apj] {10.3847/0004-637X/828/1/46}, \href
  {https://ui.adsabs.harvard.edu/abs/2016ApJ...828...46A} {828, 46}

\bibitem[\protect\citeauthoryear{{Ansdell}, {Williams}, {Manara}, {Miotello},
  {Facchini}, {van der Marel}, {Testi}  \& {van Dishoeck}}{{Ansdell}
  et~al.}{2017}]{2017AJ....153..240A}
{Ansdell} M.,  {Williams} J.~P.,  {Manara} C.~F.,  {Miotello} A.,  {Facchini}
  S.,  {van der Marel} N.,  {Testi} L.,   {van Dishoeck} E.~F.,  2017, \mn@doi
  [\aj] {10.3847/1538-3881/aa69c0}, \href
  {http://adsabs.harvard.edu/abs/2017AJ....153..240A} {153, 240}

\bibitem[\protect\citeauthoryear{{Arce} et~al.,}{{Arce}
  et~al.}{2016}]{2016hst..prop14624A}
{Arce} H.~G.,  et~al., 2016, {Taming the Flame: A Near-IR imaging study of the
  NGC 2024 (Flame Nebula) cluster}, HST Proposal

\bibitem[\protect\citeauthoryear{{Astropy Collaboration} et~al.,}{{Astropy
  Collaboration} et~al.}{2013}]{astropy:2013}
{Astropy Collaboration} et~al., 2013, \mn@doi [\aap]
  {10.1051/0004-6361/201322068}, \href
  {http://adsabs.harvard.edu/abs/2013A%26A...558A..33A} {558, A33}

\bibitem[\protect\citeauthoryear{{Bailer-Jones}, {Rybizki}, {Fouesneau},
  {Mantelet}  \& {Andrae}}{{Bailer-Jones} et~al.}{2018}]{2018AJ....156...58B}
{Bailer-Jones} C.~A.~L.,  {Rybizki} J.,  {Fouesneau} M.,  {Mantelet} G.,
  {Andrae} R.,  2018, \mn@doi [\aj] {10.3847/1538-3881/aacb21}, \href
  {https://ui.adsabs.harvard.edu/abs/2018AJ....156...58B} {156, 58}

\bibitem[\protect\citeauthoryear{{Bally}, {O'Dell}  \& {McCaughrean}}{{Bally}
  et~al.}{2000}]{2000AJ....119.2919B}
{Bally} J.,  {O'Dell} C.~R.,   {McCaughrean} M.~J.,  2000, \mn@doi [\aj]
  {10.1086/301385}, \href
  {https://ui.adsabs.harvard.edu/abs/2000AJ....119.2919B} {119, 2919}

\bibitem[\protect\citeauthoryear{{Bally}, {Youngblood}  \& {Ginsburg}}{{Bally}
  et~al.}{2012}]{2012ApJ...756..137B}
{Bally} J.,  {Youngblood} A.,   {Ginsburg} A.,  2012, \mn@doi [\apj]
  {10.1088/0004-637X/756/2/137}, \href
  {https://ui.adsabs.harvard.edu/abs/2012ApJ...756..137B} {756, 137}

\bibitem[\protect\citeauthoryear{{Barnes}, {Crutcher}, {Bieging}, {Storey}  \&
  {Willner}}{{Barnes} et~al.}{1989}]{1989ApJ...342..883B}
{Barnes} P.~J.,  {Crutcher} R.~M.,  {Bieging} J.~H.,  {Storey} J.~W.~V.,
  {Willner} S.~P.,  1989, \mn@doi [\apj] {10.1086/167645}, \href
  {https://ui.adsabs.harvard.edu/abs/1989ApJ...342..883B} {342, 883}

\bibitem[\protect\citeauthoryear{{Bik}, {Lenorzer}, {Kaper}, {Comer{\'o}n},
  {Waters}, {de Koter}  \& {Hanson}}{{Bik} et~al.}{2003}]{2003A&A...404..249B}
{Bik} A.,  {Lenorzer} A.,  {Kaper} L.,  {Comer{\'o}n} F.,  {Waters}
  L.~B.~F.~M.,  {de Koter} A.,   {Hanson} M.~M.,  2003, \mn@doi [\aap]
  {10.1051/0004-6361:20030301}, \href
  {https://ui.adsabs.harvard.edu/abs/2003A&A...404..249B} {404, 249}

\bibitem[\protect\citeauthoryear{{Birnstiel}, {Klahr}  \&
  {Ercolano}}{{Birnstiel} et~al.}{2012}]{2012A&A...539A.148B}
{Birnstiel} T.,  {Klahr} H.,   {Ercolano} B.,  2012, \mn@doi [\aap]
  {10.1051/0004-6361/201118136}, \href
  {https://ui.adsabs.harvard.edu/abs/2012A&A...539A.148B} {539, A148}

\bibitem[\protect\citeauthoryear{{Burgh}, {France}  \& {Snow}}{{Burgh}
  et~al.}{2012}]{2012ApJ...756L...6B}
{Burgh} E.~B.,  {France} K.,   {Snow} T.~P.,  2012, \mn@doi [\apjl]
  {10.1088/2041-8205/756/1/L6}, \href
  {https://ui.adsabs.harvard.edu/abs/2012ApJ...756L...6B} {756, L6}

\bibitem[\protect\citeauthoryear{{Cabrit}, {Pety}, {Pesenti}  \&
  {Dougados}}{{Cabrit} et~al.}{2006}]{2006A&A...452..897C}
{Cabrit} S.,  {Pety} J.,  {Pesenti} N.,   {Dougados} C.,  2006, \mn@doi [\aap]
  {10.1051/0004-6361:20054047}, \href
  {http://adsabs.harvard.edu/abs/2006A%26A...452..897C} {452, 897}

\bibitem[\protect\citeauthoryear{{Cadman}, {Rice}, {Hall}, {Haworth}  \&
  {Biller}}{{Cadman} et~al.}{2020}]{2020MNRAS.492.5041C}
{Cadman} J.,  {Rice} K.,  {Hall} C.,  {Haworth} T.~J.,   {Biller} B.,  2020,
  \mn@doi [\mnras] {10.1093/mnras/staa187}, \href
  {https://ui.adsabs.harvard.edu/abs/2020MNRAS.492.5041C} {492, 5041}

\bibitem[\protect\citeauthoryear{{Close} \& {Pittard}}{{Close} \&
  {Pittard}}{2017}]{2017MNRAS.469.1117C}
{Close} J.~L.,  {Pittard} J.~M.,  2017, \mn@doi [\mnras]
  {10.1093/mnras/stx897}, \href
  {https://ui.adsabs.harvard.edu/abs/2017MNRAS.469.1117C} {469, 1117}

\bibitem[\protect\citeauthoryear{{Concha-Ram{\'\i}rez}, {Vaher}  \& {Portegies
  Zwart}}{{Concha-Ram{\'\i}rez} et~al.}{2019a}]{2019MNRAS.482..732C}
{Concha-Ram{\'\i}rez} F.,  {Vaher} E.,   {Portegies Zwart} S.,  2019a, \mn@doi
  [\mnras] {10.1093/mnras/sty2721}, \href
  {https://ui.adsabs.harvard.edu/abs/2019MNRAS.482..732C} {482, 732}

\bibitem[\protect\citeauthoryear{{Concha-Ram{\'\i}rez}, {Wilhelm}, {Portegies
  Zwart}  \& {Haworth}}{{Concha-Ram{\'\i}rez}
  et~al.}{2019b}]{2019MNRAS.490.5678C}
{Concha-Ram{\'\i}rez} F.,  {Wilhelm} M. J.~C.,  {Portegies Zwart} S.,
  {Haworth} T.~J.,  2019b, \mn@doi [\mnras] {10.1093/mnras/stz2973}, \href
  {https://ui.adsabs.harvard.edu/abs/2019MNRAS.490.5678C} {490, 5678}

\bibitem[\protect\citeauthoryear{{Dai}, {Facchini}, {Clarke}  \&
  {Haworth}}{{Dai} et~al.}{2015}]{2015MNRAS.449.1996D}
{Dai} F.,  {Facchini} S.,  {Clarke} C.~J.,   {Haworth} T.~J.,  2015, \mn@doi
  [\mnras] {10.1093/mnras/stv403}, \href
  {http://adsabs.harvard.edu/abs/2015MNRAS.449.1996D} {449, 1996}

\bibitem[\protect\citeauthoryear{{Eisner} et~al.,}{{Eisner}
  et~al.}{2018}]{2018ApJ...860...77E}
{Eisner} J.~A.,  et~al., 2018, \mn@doi [\apj] {10.3847/1538-4357/aac3e2}, \href
  {http://adsabs.harvard.edu/abs/2018ApJ...860...77E} {860, 77}

\bibitem[\protect\citeauthoryear{{Facchini}, {Clarke}  \& {Bisbas}}{{Facchini}
  et~al.}{2016}]{2016MNRAS.457.3593F}
{Facchini} S.,  {Clarke} C.~J.,   {Bisbas} T.~G.,  2016, \mn@doi [\mnras]
  {10.1093/mnras/stw240}, \href
  {http://adsabs.harvard.edu/abs/2016MNRAS.457.3593F} {457, 3593}

\bibitem[\protect\citeauthoryear{{Fang} et~al.,}{{Fang}
  et~al.}{2012}]{2012A&A...539A.119F}
{Fang} M.,  et~al., 2012, \mn@doi [\aap] {10.1051/0004-6361/201015914}, \href
  {https://ui.adsabs.harvard.edu/abs/2012A&A...539A.119F} {539, A119}

\bibitem[\protect\citeauthoryear{{Garrison}}{{Garrison}}{1968}]{1968PASP...80...20G}
{Garrison} R.~F.,  1968, \mn@doi [\pasp] {10.1086/128580}, \href
  {https://ui.adsabs.harvard.edu/abs/1968PASP...80...20G} {80, 20}

\bibitem[\protect\citeauthoryear{{Getman}, {Feigelson}  \& {Kuhn}}{{Getman}
  et~al.}{2014}]{2014ApJ...787..109G}
{Getman} K.~V.,  {Feigelson} E.~D.,   {Kuhn} M.~A.,  2014, \mn@doi [\apj]
  {10.1088/0004-637X/787/2/109}, \href
  {https://ui.adsabs.harvard.edu/abs/2014ApJ...787..109G} {787, 109}

\bibitem[\protect\citeauthoryear{{Graham}, {Meaburn}, {Garrington}, {O'Brien},
  {Henney}  \& {O'Dell}}{{Graham} et~al.}{2002}]{2002ApJ...570..222G}
{Graham} M.~F.,  {Meaburn} J.,  {Garrington} S.~T.,  {O'Brien} T.~J.,  {Henney}
  W.~J.,   {O'Dell} C.~R.,  2002, \mn@doi [\apj] {10.1086/339398}, \href
  {https://ui.adsabs.harvard.edu/abs/2002ApJ...570..222G} {570, 222}

\bibitem[\protect\citeauthoryear{{Guarcello} et~al.,}{{Guarcello}
  et~al.}{2016}]{2016arXiv160501773G}
{Guarcello} M.~G.,  et~al., 2016, arXiv e-prints, \href
  {https://ui.adsabs.harvard.edu/abs/2016arXiv160501773G} {p. arXiv:1605.01773}

\bibitem[\protect\citeauthoryear{{Haisch}, {Lada}  \& {Lada}}{{Haisch}
  et~al.}{2000}]{2000AJ....120.1396H}
{Haisch} Karl~E. J.,  {Lada} E.~A.,   {Lada} C.~J.,  2000, \mn@doi [\aj]
  {10.1086/301521}, \href
  {https://ui.adsabs.harvard.edu/abs/2000AJ....120.1396H} {120, 1396}

\bibitem[\protect\citeauthoryear{{Haisch}, {Lada}, {Pi{\~n}a}, {Telesco}  \&
  {Lada}}{{Haisch} et~al.}{2001a}]{2001AJ....121.1512H}
{Haisch} Karl~E. J.,  {Lada} E.~A.,  {Pi{\~n}a} R.~K.,  {Telesco} C.~M.,
  {Lada} C.~J.,  2001a, \mn@doi [\aj] {10.1086/319397}, \href
  {https://ui.adsabs.harvard.edu/abs/2001AJ....121.1512H} {121, 1512}

\bibitem[\protect\citeauthoryear{{Haisch}, {Lada}  \& {Lada}}{{Haisch}
  et~al.}{2001b}]{2001ApJ...553L.153H}
{Haisch} Jr. K.~E.,  {Lada} E.~A.,   {Lada} C.~J.,  2001b, \mn@doi [\apjl]
  {10.1086/320685}, \href {http://adsabs.harvard.edu/abs/2001ApJ...553L.153H}
  {553, L153}

\bibitem[\protect\citeauthoryear{{Haworth} \& {Clarke}}{{Haworth} \&
  {Clarke}}{2019}]{2019MNRAS.485.3895H}
{Haworth} T.~J.,  {Clarke} C.~J.,  2019, \mn@doi [\mnras]
  {10.1093/mnras/stz706}, \href
  {https://ui.adsabs.harvard.edu/abs/2019MNRAS.485.3895H} {485, 3895}

\bibitem[\protect\citeauthoryear{{Haworth} \& {Owen}}{{Haworth} \&
  {Owen}}{2020}]{2020MNRAS.492.5030H}
{Haworth} T.~J.,  {Owen} J.~E.,  2020, \mn@doi [\mnras]
  {10.1093/mnras/staa151}, \href
  {https://ui.adsabs.harvard.edu/abs/2020MNRAS.492.5030H} {492, 5030}

\bibitem[\protect\citeauthoryear{{Haworth}, {Facchini}, {Clarke}  \&
  {Cleeves}}{{Haworth} et~al.}{2017}]{2017MNRAS.468L.108H}
{Haworth} T.~J.,  {Facchini} S.,  {Clarke} C.~J.,   {Cleeves} L.~I.,  2017,
  \mn@doi [\mnras] {10.1093/mnrasl/slx037}, \href
  {http://adsabs.harvard.edu/abs/2017MNRAS.468L.108H} {468, L108}

\bibitem[\protect\citeauthoryear{{Haworth}, {Facchini}, {Clarke}  \&
  {Mohanty}}{{Haworth} et~al.}{2018a}]{2018MNRAS.475.5460H}
{Haworth} T.~J.,  {Facchini} S.,  {Clarke} C.~J.,   {Mohanty} S.,  2018a,
  \mn@doi [\mnras] {10.1093/mnras/sty168}, \href
  {https://ui.adsabs.harvard.edu/abs/2018MNRAS.475.5460H} {475, 5460}

\bibitem[\protect\citeauthoryear{{Haworth}, {Clarke}, {Rahman}, {Winter}  \&
  {Facchini}}{{Haworth} et~al.}{2018b}]{2018MNRAS.481..452H}
{Haworth} T.~J.,  {Clarke} C.~J.,  {Rahman} W.,  {Winter} A.~J.,   {Facchini}
  S.,  2018b, \mn@doi [\mnras] {10.1093/mnras/sty2323}, \href
  {http://adsabs.harvard.edu/abs/2018MNRAS.481..452H} {481, 452}

\bibitem[\protect\citeauthoryear{{Haworth}, {Cadman}, {Meru}, {Hall},
  {Albertini}, {Forgan}, {Rice}  \& {Owen}}{{Haworth}
  et~al.}{2020}]{2020MNRAS.494.4130H}
{Haworth} T.~J.,  {Cadman} J.,  {Meru} F.,  {Hall} C.,  {Albertini} E.,
  {Forgan} D.,  {Rice} K.,   {Owen} J.~E.,  2020, \mn@doi [\mnras]
  {10.1093/mnras/staa883}, \href
  {https://ui.adsabs.harvard.edu/abs/2020MNRAS.494.4130H} {494, 4130}

\bibitem[\protect\citeauthoryear{{Henney}}{{Henney}}{2002}]{2002RMxAA..38...71H}
{Henney} W.~J.,  2002, \rmxaa, \href
  {https://ui.adsabs.harvard.edu/abs/2002RMxAA..38...71H} {38, 71}

\bibitem[\protect\citeauthoryear{{Henney} \& {O'Dell}}{{Henney} \&
  {O'Dell}}{1999}]{1999AJ....118.2350H}
{Henney} W.~J.,  {O'Dell} C.~R.,  1999, \mn@doi [\aj] {10.1086/301087}, \href
  {https://ui.adsabs.harvard.edu/abs/1999AJ....118.2350H} {118, 2350}

\bibitem[\protect\citeauthoryear{{Henney}, {O'Dell}, {Meaburn}, {Garrington}
  \& {Lopez}}{{Henney} et~al.}{2002}]{2002ApJ...566..315H}
{Henney} W.~J.,  {O'Dell} C.~R.,  {Meaburn} J.,  {Garrington} S.~T.,   {Lopez}
  J.~A.,  2002, \mn@doi [\apj] {10.1086/338055}, \href
  {http://adsabs.harvard.edu/abs/2002ApJ...566..315H} {566, 315}

\bibitem[\protect\citeauthoryear{{Hodapp}, {Iserlohe}, {Stecklum}  \&
  {Krabbe}}{{Hodapp} et~al.}{2009}]{2009ApJ...701L.100H}
{Hodapp} K.~W.,  {Iserlohe} C.,  {Stecklum} B.,   {Krabbe} A.,  2009, \mn@doi
  [\apjl] {10.1088/0004-637X/701/2/L100}, \href
  {https://ui.adsabs.harvard.edu/abs/2009ApJ...701L.100H} {701, L100}

\bibitem[\protect\citeauthoryear{{Johnstone}, {Hollenbach}  \&
  {Bally}}{{Johnstone} et~al.}{1998}]{Johnstone98}
{Johnstone} D.,  {Hollenbach} D.,   {Bally} J.,  1998, \mn@doi [\apj]
  {10.1086/305658}, \href
  {https://ui.adsabs.harvard.edu/abs/1998ApJ...499..758J} {499, 758}

\bibitem[\protect\citeauthoryear{{Kandori} et~al.,}{{Kandori}
  et~al.}{2007}]{2007PASJ...59..487K}
{Kandori} R.,  et~al., 2007, \mn@doi [\pasj] {10.1093/pasj/59.3.487}, \href
  {https://ui.adsabs.harvard.edu/abs/2007PASJ...59..487K} {59, 487}

\bibitem[\protect\citeauthoryear{{Kim}, {Clarke}, {Fang}  \& {Facchini}}{{Kim}
  et~al.}{2016}]{Kim16}
{Kim} J.~S.,  {Clarke} C.~J.,  {Fang} M.,   {Facchini} S.,  2016, \mn@doi
  [\apjl] {10.3847/2041-8205/826/1/L15}, \href
  {http://adsabs.harvard.edu/abs/2016ApJ...826L..15K} {826, L15}

\bibitem[\protect\citeauthoryear{{Kounkel} et~al.,}{{Kounkel}
  et~al.}{2018}]{2018AJ....156...84K}
{Kounkel} M.,  et~al., 2018, \mn@doi [\aj] {10.3847/1538-3881/aad1f1}, \href
  {https://ui.adsabs.harvard.edu/abs/2018AJ....156...84K} {156, 84}

\bibitem[\protect\citeauthoryear{{Kruegel}, {Thum}, {Pankonin}  \&
  {Martin-Pintado}}{{Kruegel} et~al.}{1982}]{1982A&AS...48..345K}
{Kruegel} E.,  {Thum} C.,  {Pankonin} V.,   {Martin-Pintado} J.,  1982, \aaps,
  \href {https://ui.adsabs.harvard.edu/abs/1982A&AS...48..345K} {48, 345}

\bibitem[\protect\citeauthoryear{{Kuffmeier}, {Frostholm Mogensen},
  {Haugb{\o}lle}, {Bizzarro}  \& {Nordlund}}{{Kuffmeier}
  et~al.}{2016}]{2016ApJ...826...22K}
{Kuffmeier} M.,  {Frostholm Mogensen} T.,  {Haugb{\o}lle} T.,  {Bizzarro} M.,
  {Nordlund} {\r{A}}.,  2016, \mn@doi [\apj] {10.3847/0004-637X/826/1/22},
  \href {https://ui.adsabs.harvard.edu/abs/2016ApJ...826...22K} {826, 22}

\bibitem[\protect\citeauthoryear{{Kuffmeier}, {Zhao}  \& {Caselli}}{{Kuffmeier}
  et~al.}{2020}]{2020arXiv200600019K}
{Kuffmeier} M.,  {Zhao} B.,   {Caselli} P.,  2020, arXiv e-prints, \href
  {https://ui.adsabs.harvard.edu/abs/2020arXiv200600019K} {p. arXiv:2006.00019}

\bibitem[\protect\citeauthoryear{{Levine}, {Steinhauer}, {Elston}  \&
  {Lada}}{{Levine} et~al.}{2006}]{2006ApJ...646.1215L}
{Levine} J.~L.,  {Steinhauer} A.,  {Elston} R.~J.,   {Lada} E.~A.,  2006,
  \mn@doi [\apj] {10.1086/504964}, \href
  {https://ui.adsabs.harvard.edu/abs/2006ApJ...646.1215L} {646, 1215}

\bibitem[\protect\citeauthoryear{{Lichtenberg}, {Golabek}, {Burn}, {Meyer},
  {Alibert}, {Gerya}  \& {Mordasini}}{{Lichtenberg}
  et~al.}{2019}]{2019NatAs...3..307L}
{Lichtenberg} T.,  {Golabek} G.~J.,  {Burn} R.,  {Meyer} M.~R.,  {Alibert} Y.,
  {Gerya} T.~V.,   {Mordasini} C.,  2019, \mn@doi [Nature Astronomy]
  {10.1038/s41550-018-0688-5}, \href
  {https://ui.adsabs.harvard.edu/abs/2019NatAs...3..307L} {3, 307}

\bibitem[\protect\citeauthoryear{{Mann} et~al.,}{{Mann}
  et~al.}{2014}]{2014ApJ...784...82M}
{Mann} R.~K.,  et~al., 2014, \mn@doi [\apj] {10.1088/0004-637X/784/1/82}, \href
  {https://ui.adsabs.harvard.edu/abs/2014ApJ...784...82M} {784, 82}

\bibitem[\protect\citeauthoryear{{Marcy} et~al.,}{{Marcy}
  et~al.}{2014}]{2014ApJS..210...20M}
{Marcy} G.~W.,  et~al., 2014, \mn@doi [\apjs] {10.1088/0067-0049/210/2/20},
  \href {https://ui.adsabs.harvard.edu/abs/2014ApJS..210...20M} {210, 20}

\bibitem[\protect\citeauthoryear{{Mauc{\'o}} et~al.,}{{Mauc{\'o}}
  et~al.}{2016}]{2016ApJ...829...38M}
{Mauc{\'o}} K.,  et~al., 2016, \mn@doi [\apj] {10.3847/0004-637X/829/1/38},
  \href {https://ui.adsabs.harvard.edu/abs/2016ApJ...829...38M} {829, 38}

\bibitem[\protect\citeauthoryear{{McCaughrean} \& {O'dell}}{{McCaughrean} \&
  {O'dell}}{1996}]{1996AJ....111.1977M}
{McCaughrean} M.~J.,  {O'dell} C.~R.,  1996, \mn@doi [\aj] {10.1086/117934},
  \href {http://adsabs.harvard.edu/abs/1996AJ....111.1977M} {111, 1977}

\bibitem[\protect\citeauthoryear{{Menten}, {Reid}, {Forbrich}  \&
  {Brunthaler}}{{Menten} et~al.}{2007}]{2007A&A...474..515M}
{Menten} K.~M.,  {Reid} M.~J.,  {Forbrich} J.,   {Brunthaler} A.,  2007,
  \mn@doi [\aap] {10.1051/0004-6361:20078247}, \href
  {https://ui.adsabs.harvard.edu/abs/2007A&A...474..515M} {474, 515}

\bibitem[\protect\citeauthoryear{{Meyer}}{{Meyer}}{1996}]{1996PhDT.........3M}
{Meyer} M.~R.,  1996, PhD thesis, Max-Planck-Institut f{\"u}r Astronomie,
  K{\"o}nigstuhl 17, D-69117 Heidelberg, Germany

\bibitem[\protect\citeauthoryear{{Moore}, {Li}  \& {Adams}}{{Moore}
  et~al.}{2020}]{2020arXiv200715666M}
{Moore} N. W.~H.,  {Li} G.,   {Adams} F.~C.,  2020, arXiv e-prints, \href
  {https://ui.adsabs.harvard.edu/abs/2020arXiv200715666M} {p. arXiv:2007.15666}

\bibitem[\protect\citeauthoryear{{Nicholson}, {Parker}, {Church}, {Davies},
  {Fearon}  \& {Walton}}{{Nicholson} et~al.}{2019}]{2019MNRAS.485.4893N}
{Nicholson} R.~B.,  {Parker} R.~J.,  {Church} R.~P.,  {Davies} M.~B.,  {Fearon}
  N.~M.,   {Walton} S. R.~J.,  2019, \mn@doi [\mnras] {10.1093/mnras/stz606},
  \href {https://ui.adsabs.harvard.edu/abs/2019MNRAS.485.4893N} {485, 4893}

\bibitem[\protect\citeauthoryear{{O'dell}}{{O'dell}}{1998}]{1998AJ....115..263O}
{O'dell} C.~R.,  1998, \mn@doi [\aj] {10.1086/300178}, \href
  {http://adsabs.harvard.edu/abs/1998AJ....115..263O} {115, 263}

\bibitem[\protect\citeauthoryear{{O'dell} \& {Wen}}{{O'dell} \&
  {Wen}}{1994}]{1994ApJ...436..194O}
{O'dell} C.~R.,  {Wen} Z.,  1994, \mn@doi [\apj] {10.1086/174892}, \href
  {https://ui.adsabs.harvard.edu/abs/1994ApJ...436..194O} {436, 194}

\bibitem[\protect\citeauthoryear{{Ormel}, {Liu}  \& {Schoonenberg}}{{Ormel}
  et~al.}{2017}]{Ormel17Trappist}
{Ormel} C.~W.,  {Liu} B.,   {Schoonenberg} D.,  2017, \mn@doi [\aap]
  {10.1051/0004-6361/201730826}, \href
  {http://adsabs.harvard.edu/abs/2017A%26A...604A...1O} {604, A1}

\bibitem[\protect\citeauthoryear{{Parker}}{{Parker}}{2020}]{2020arXiv200707890P}
{Parker} R.~J.,  2020, arXiv e-prints, \href
  {https://ui.adsabs.harvard.edu/abs/2020arXiv200707890P} {p. arXiv:2007.07890}

\bibitem[\protect\citeauthoryear{{Price-Whelan} et~al.,}{{Price-Whelan}
  et~al.}{2018}]{astropy:2018}
{Price-Whelan} A.~M.,  et~al., 2018, \mn@doi [\aj] {10.3847/1538-3881/aabc4f},
  \href {https://ui.adsabs.harvard.edu/#abs/2018AJ....156..123T} {156, 123}

\bibitem[\protect\citeauthoryear{{Reiter} \& {Parker}}{{Reiter} \&
  {Parker}}{2019}]{2019MNRAS.486.4354R}
{Reiter} M.,  {Parker} R.~J.,  2019, \mn@doi [\mnras] {10.1093/mnras/stz1115},
  \href {https://ui.adsabs.harvard.edu/abs/2019MNRAS.486.4354R} {486, 4354}

\bibitem[\protect\citeauthoryear{{Rodriguez} et~al.,}{{Rodriguez}
  et~al.}{2018}]{2018ApJ...859..150R}
{Rodriguez} J.~E.,  et~al., 2018, \mn@doi [\apj] {10.3847/1538-4357/aac08f},
  \href {https://ui.adsabs.harvard.edu/abs/2018ApJ...859..150R} {859, 150}

\bibitem[\protect\citeauthoryear{{Scally} \& {Clarke}}{{Scally} \&
  {Clarke}}{2001}]{2001MNRAS.325..449S}
{Scally} A.,  {Clarke} C.,  2001, \mn@doi [\mnras]
  {10.1046/j.1365-8711.2001.04274.x}, \href
  {http://adsabs.harvard.edu/abs/2001MNRAS.325..449S} {325, 449}

\bibitem[\protect\citeauthoryear{{Segura-Cox} et~al.,}{{Segura-Cox}
  et~al.}{2020}]{2020Natur.586..228S}
{Segura-Cox} D.~M.,  et~al., 2020, \mn@doi [\nat] {10.1038/s41586-020-2779-6},
  \href {https://ui.adsabs.harvard.edu/abs/2020Natur.586..228S} {586, 228}

\bibitem[\protect\citeauthoryear{{Sellek}, {Booth}  \& {Clarke}}{{Sellek}
  et~al.}{2020}]{2020MNRAS.492.1279S}
{Sellek} A.~D.,  {Booth} R.~A.,   {Clarke} C.~J.,  2020, \mn@doi [\mnras]
  {10.1093/mnras/stz3528}, \href
  {https://ui.adsabs.harvard.edu/abs/2020MNRAS.492.1279S} {492, 1279}

\bibitem[\protect\citeauthoryear{{Smith}, {Bally}  \& {Walborn}}{{Smith}
  et~al.}{2010}]{2010MNRAS.405.1153S}
{Smith} N.,  {Bally} J.,   {Walborn} N.~R.,  2010, \mn@doi [\mnras]
  {10.1111/j.1365-2966.2010.16520.x}, \href
  {https://ui.adsabs.harvard.edu/abs/2010MNRAS.405.1153S} {405, 1153}

\bibitem[\protect\citeauthoryear{{Stapelfeldt}, {Sahai}, {Werner}  \&
  {Trauger}}{{Stapelfeldt} et~al.}{1997}]{1997ASPC..119..131S}
{Stapelfeldt} K.,  {Sahai} R.,  {Werner} M.,   {Trauger} J.,  1997, in
  {Soderblom} D.,  ed.,  Astronomical Society of the Pacific Conference Series
  Vol. 119, Planets Beyond the Solar System and the Next Generation of Space
  Missions. p.~131

\bibitem[\protect\citeauthoryear{{Sternberg}, {Hoffmann}  \&
  {Pauldrach}}{{Sternberg} et~al.}{2003}]{2003ApJ...599.1333S}
{Sternberg} A.,  {Hoffmann} T.~L.,   {Pauldrach} A.~W.~A.,  2003, \mn@doi
  [\apj] {10.1086/379506}, \href
  {https://ui.adsabs.harvard.edu/abs/2003ApJ...599.1333S} {599, 1333}

\bibitem[\protect\citeauthoryear{{St{\"o}rzer} \& {Hollenbach}}{{St{\"o}rzer}
  \& {Hollenbach}}{1999}]{1999ApJ...515..669S}
{St{\"o}rzer} H.,  {Hollenbach} D.,  1999, \mn@doi [\apj] {10.1086/307055},
  \href {http://adsabs.harvard.edu/abs/1999ApJ...515..669S} {515, 669}

\bibitem[\protect\citeauthoryear{{Stutz} et~al.,}{{Stutz}
  et~al.}{2013}]{2013ApJ...767...36S}
{Stutz} A.~M.,  et~al., 2013, \mn@doi [\apj] {10.1088/0004-637X/767/1/36},
  \href {https://ui.adsabs.harvard.edu/abs/2013ApJ...767...36S} {767, 36}

\bibitem[\protect\citeauthoryear{{Thompson}, {Thronson}  \&
  {Campbell}}{{Thompson} et~al.}{1981}]{1981ApJ...249..622T}
{Thompson} R.~I.,  {Thronson} H.~A. J.,   {Campbell} B.~G.,  1981, \mn@doi
  [\apj] {10.1086/159322}, \href
  {https://ui.adsabs.harvard.edu/abs/1981ApJ...249..622T} {249, 622}

\bibitem[\protect\citeauthoryear{{Vasconcelos}, {Cerqueira}, {Raga}  \&
  {Amorim}}{{Vasconcelos} et~al.}{2010}]{2010RMxAA..46...79V}
{Vasconcelos} M.~J.,  {Cerqueira} A.~H.,  {Raga} A.~C.,   {Amorim} R.~R.,
  2010, \rmxaa, \href {https://ui.adsabs.harvard.edu/abs/2010RMxAA..46...79V}
  {46, 79}

\bibitem[\protect\citeauthoryear{{Winter}, {Clarke}, {Rosotti}, {Ih},
  {Facchini}  \& {Haworth}}{{Winter} et~al.}{2018a}]{Winter18b}
{Winter} A.~J.,  {Clarke} C.~J.,  {Rosotti} G.,  {Ih} J.,  {Facchini} S.,
  {Haworth} T.~J.,  2018a, \mn@doi [\mnras] {10.1093/mnras/sty984}, \href
  {http://adsabs.harvard.edu/abs/2018MNRAS.478.2700W} {478, 2700}

\bibitem[\protect\citeauthoryear{{Winter}, {Booth}  \& {Clarke}}{{Winter}
  et~al.}{2018b}]{Winter18c}
{Winter} A.~J.,  {Booth} R.~A.,   {Clarke} C.~J.,  2018b, \mn@doi [\mnras]
  {10.1093/mnras/sty1866}, \href
  {https://ui.adsabs.harvard.edu/abs/2018MNRAS.479.5522W} {479, 5522}

\bibitem[\protect\citeauthoryear{{Winter}, {Clarke}, {Rosotti}, {Hacar}  \&
  {Alexander}}{{Winter} et~al.}{2019}]{2019MNRAS.490.5478W}
{Winter} A.~J.,  {Clarke} C.~J.,  {Rosotti} G.~P.,  {Hacar} A.,   {Alexander}
  R.,  2019, \mn@doi [\mnras] {10.1093/mnras/stz2545}, \href
  {https://ui.adsabs.harvard.edu/abs/2019MNRAS.490.5478W} {490, 5478}

\bibitem[\protect\citeauthoryear{{Winter}, {Kruijssen}, {Chevance}, {Keller}
  \& {Longmore}}{{Winter} et~al.}{2020a}]{2020MNRAS.491..903W}
{Winter} A.~J.,  {Kruijssen} J.~M.~D.,  {Chevance} M.,  {Keller} B.~W.,
  {Longmore} S.~N.,  2020a, \mn@doi [\mnras] {10.1093/mnras/stz2747}, \href
  {https://ui.adsabs.harvard.edu/abs/2020MNRAS.491..903W} {491, 903}

\bibitem[\protect\citeauthoryear{{Winter}, {Ansdell}, {Haworth}  \&
  {Kruijssen}}{{Winter} et~al.}{2020b}]{2020MNRAS.497L..40W}
{Winter} A.~J.,  {Ansdell} M.,  {Haworth} T.~J.,   {Kruijssen} J.~M.~D.,
  2020b, \mn@doi [\mnras] {10.1093/mnrasl/slaa110}, \href
  {https://ui.adsabs.harvard.edu/abs/2020MNRAS.497L..40W} {497, L40}

\bibitem[\protect\citeauthoryear{{Wu} et~al.,}{{Wu}
  et~al.}{2013}]{2013ApJ...774...45W}
{Wu} Y.~L.,  et~al., 2013, \mn@doi [\apj] {10.1088/0004-637X/774/1/45}, \href
  {https://ui.adsabs.harvard.edu/abs/2013ApJ...774...45W} {774, 45}

\bibitem[\protect\citeauthoryear{{Wu}, {Close}, {Kim}, {Males}  \&
  {Morzinski}}{{Wu} et~al.}{2018}]{2018ApJ...854..144W}
{Wu} Y.-L.,  {Close} L.~M.,  {Kim} J.~S.,  {Males} J.~R.,   {Morzinski} K.~M.,
  2018, \mn@doi [\apj] {10.3847/1538-4357/aaa96b}, \href
  {https://ui.adsabs.harvard.edu/abs/2018ApJ...854..144W} {854, 144}

\bibitem[\protect\citeauthoryear{{van Terwisga} et~al.,}{{van Terwisga}
  et~al.}{2020}]{vanTerwisga20}
{van Terwisga} S.~E.,  et~al., 2020, \mn@doi [\aap]
  {10.1051/0004-6361/201937403}, \href
  {https://ui.adsabs.harvard.edu/abs/2020A&A...640A..27V} {640, A27}

\makeatother
\end{thebibliography}

\end{document}